\documentclass[11pt]{article}
\usepackage{hyperref}
\usepackage{color}
\usepackage{amsmath, amsfonts, amsthm, amssymb, graphicx}
\usepackage{xspace}
\usepackage{epsfig}
\usepackage{psfrag}

\newcommand{\Real}{\ensuremath{\mathbb{R}}\xspace}

\newcommand{\op}{\ensuremath{\mathcal{O}}\xspace}
\newcommand{\del}{\partial} 
\newcommand{\cdel}{\nabla}
\newcommand{\fdel}[2][]{\ensuremath{\frac{\delta #1}{\delta #2}}}

\newcommand{\vev}[1]{\ensuremath{\langle #1 \rangle}\xspace}

\def\ie{{\it i.e.\ }}

\newcommand{\hf}{\frac{1}{2}}
\newcommand{\qt}{\frac{1}{4}}

\let\a=\alpha  \let\g=\gamma \let\d=\delta 
    \let\k=\kappa
 \let\m=\mu \let\n=\nu \let\x=\xi \let\r=\rho
\let\s=\sigma     
  \let\D=\Delta  
    
    \let\G=\Gamma

\newcommand{\be}{\begin{equation}}
\newcommand{\ee}{\end{equation}}
\def\ba{\begin{array}}
\def\ea{\end{array}}
\newcommand{\bea}{\begin{eqnarray}}
\newcommand{\eea}{\end{eqnarray}}

\newcommand{\as}[1]{_{\{ #1 \}}}
\newcommand{\asct}[1]{_{ #1 ,\text{ct}}}
\newcommand{\asctexp}[2]{_{ #1 ,\text{ct}\,\{ #2 \}}}
\newcommand{\bs}[1]{_{( #1 )}}
\newcommand{\ab}[2]{_{\{ #1 \} (#2) }}
\newcommand{\Dp}{\Delta_+}
\newcommand{\Dm}{\Delta_-}
\newcommand{\Dpm}{\Delta_\pm}

\begin{document}
\title{Holographic renormalization for irrelevant operators \\ and multi-trace counterterms}
\author{Balt C. van Rees\thanks{vanrees@insti.physics.sunysb.edu}}
\date{}
\maketitle

\begin{center}
\it C. N. Yang Institute for Theoretical Physics\\
State University of New York, Stony Brook, NY 11794-3840
\end{center}

\vskip 1cm
\begin{abstract}
We investigate the structure of holographic renormalization in the presence of sources for irrelevant operators. By working perturbatively in the sources we avoid issues related to the non-renormalizability of the dual field theory. We find new classes of divergences which appear to be non-local on the gravity side. However in all cases a systematic renormalization procedure exists involving either standard local counterterms or new counterterms which may be interpreted as multi-trace counterterms in the field theory. The multi-trace counterterms reflect a more intricate relation between sources and the asymptotics of bulk fields.
\end{abstract}

\newpage
\tableofcontents
\newpage

\section{Introduction}
In recent years a variety of conjectures has been made concerning a possible holographic interpretation of spacetimes which are not asymptotically (locally) of an AdS form. In particular, we have in mind examples like the non-relativistic spacetimes of \cite{Son:2008ye,Balasubramanian:2008dm}, the Lifshitz geometries \cite{Kachru:2008yh} but also the longer known geometries dual to non-commutative gauge theories \cite{Hashimoto:1999ut,Maldacena:1999mh}. A common feature of these spacetimes is that they all have a peculiar asymptotic structure where some of the metric components diverge faster near the boundary of the spacetime than would be the case for an Asymptotically locally AdS (or AlAdS) background. For example, in the $d+1$-dimensional non-relativistic backgrounds of \cite{Son:2008ye} the metric has the form:
\be
\label{eq:nrmetric}
ds^2 = dr^2 - b^2 e^{4r} du^2 + e^{2r} (du dv + dx^i dx^i)\,,  
\ee
where $i \in \{1,2,\ldots ,d-2\}$. When the parameter $b$ vanishes we recover the usual AdS metric, but for nonzero $b$ the $(uu)$-component of the metric diverges faster near the boundary $r \to \infty$ than the AdS components.

Such a modification of the asymptotics of the bulk metric implies the dual field theory is no longer conformal in the UV. Indeed, for the metric \eqref{eq:nrmetric} it was suggested in \cite{Son:2008ye} and worked out in more detail in \cite{us,Costa:2010cn} that the dual field theory is deformed by an \emph{irrelevant} operator. This can be seen by taking the small $b$ limit, where \eqref{eq:nrmetric} corresponds to a small perturbation around the AdS metric. We can holographically interpret this situation using the standard AdS/CFT dictionary and conclude that the deformation to first order corresponds to switching on a nonzero source $b$ for an irrelevant vector operator. In fact, such a deformation picture applies not only to all the examples mentioned above, but also seems to occur in the NHEK geometry of \cite{Guica:2008mu} as was recently discussed in \cite{Compere:2010uk,Guica:2010ej}.

The fact that the deforming operator is irrelevant leads to genuine concerns about the renormalizability of the dual field theory at finite $b$. These concerns should be reflected in the bulk theory as well, but it is currently not known to which extend the on-shell gravity action for perturbations around these backgrounds can be holographically renormalized. We recall that many of the key results of holographic renormalization (see \cite{Skenderis:2002wp} for an introduction) largely apply to AlAdS spacetimes only and cannot be directly applied to non-AlAdS metrics like \eqref{eq:nrmetric}. In order to draw any meaningful conclusions concerning the renormalizability of the on-shell action for such modified asymptotics it will therefore be necessary to extend the currently known results.

In this paper we will take a first step in this direction by considering the holographic renormalization procedure in the presence of \emph{parametrically small} sources for irrelevant operators. If we work up to any given finite order in the sources then the dual field theory remains fully renormalizable and the same result should apply to the on-shell bulk action as well. We will demonstrate below that the bulk action indeed remains renormalizable, although the usual methods of holographic renormalization need to be extended in a nontrivial manner to incorporate even such parametrically small irrelevant deformations. In the remainder of this paper we work out the detailed holographic renormalization procedure only in very specific examples, but the general conclusions we draw should be applicable in all cases where one holographically computes correlation functions involving irrelevant operators. As such the present work forms an integral part of any complete gauge/gravity dictionary. We however expect our results to be particularly useful in the study of holography for the kind of non-AlAdS spacetimes we mentioned above.

\subsection{Summary of results}
The main computations of this paper concern the holographic renormalization of an on-shell bulk action in two toy model examples. The first example, discussed in section \ref{sec:nologs}, is a massive scalar field $\Phi$ with a $\lambda \Phi^3$ interaction, propagating in a fixed $d+1$-dimensional AdS background. The second example is a free massive scalar field propagating in a general $d+1$-dimensional AlAdS spacetime and will be discussed in section \ref{sec:gravity}. Since the scalar fields are always taken to be massive, they correspond as promised to irrelevant operators. In both examples the radial expansion of $\Phi$ contains a term of the form (see the next section for our conventions):
\be
\Phi(r,x^i) = \phi_{(0)}(x^i) e^{(\D - d)r } + \ldots
\ee
In the gauge/gravity duality we interpret $\phi_{(0)}$ as the source for an operator $\op$ dual to $\Phi$. 

Our first result is the following. As is reviewed in \cite{Skenderis:2002wp}, in all the examples of holographic renormalization worked out so far, the divergences are always \emph{local} functionals of the sources of the dual field theory, so for the scalar field they would be local functionals of $\phi_{(0)}$. For irrelevant deformations however we find that this no longer has to be the case. Instead, it may occur that the divergences are \emph{non-local} functions of $\phi_{(0)}$ but nevertheless they are \emph{local} functions of the fields at the cutoff surface $r=r_0$, that is of $\Phi(r_0)$. We refer to such divergences as `pseudo-non-local', since they can be renormalized with local counterterms at the cutoff surface.

Our second result is however that not all non-local divergences can be cancelled in this way. Some divergences necessitate the use of counterterms which are genuinely non-local when expressed in terms of the field $\Phi$. They are however \emph{local} when expressed in terms of the field $\Phi$ and its conjugate momentum $\del_r \Phi$. Furthermore, in all the examples we considered they are at least quadratic in $\del_r \Phi$. We interpret these counterterms as \emph{multi-trace} counterterms in the dual field theory. The fact that such counterterms are necessary of course has repercussions on the variational principle in the bulk on which we will comment below. The presence of these multi-trace counterterms in the standard framework of holographic renormalization is again a new result, although their role has recently been emphasized in a different framework in \cite{Heemskerk:2010hk}. 
 
Our third and last result concerns the backreaction onto other fields. We find that in our second toy model computation, so the holographic renormalization for a free scalar field in a general AlAdS spacetime, it is \emph{necessary} to take into account the backreaction. A simple computation in section \ref{sec:gravity} demonstrates that the effect of the backreaction is indeed in no sense smaller than the original perturbation and that indeed non-local counterterms (which are neither pseudo-non-local nor of multi-trace form) can only be avoided upon taking into account the backreaction onto the metric. Once the backreaction has been accounted for we will find the same `pseudo-non-local' and multi-trace divergences as in our first example.

Although we have performed no further computations, we expect that the general procedure of holographic renormalization is reflected accurately by these examples. In particular, the dual field theory is renormalizable with single- and multi-trace counterterms for any parametrically small irrelevant deformations. In the bulk we therefore expect to need only counterterms which are either local functionals of the boundary fields or multi-trace counterterms which are by definition at least quadratic in the conjugate momenta. In other words, there should be no counterterms which are \emph{linear} in the conjugate momenta since these cannot be interpreted as local counterterms in the field theory. Any divergence linear in the conjugate momenta will therefore have to be pseudo-non-local and should be cancelled by other counterterms. We emphasize that this is a nontrivial conjecture concerning general perturbations around AlAdS spacetimes which generalizes the well-known results that the on-shell action for an AlAdS spacetime can always be renormalized with local counterterms. (Notice that the latter statement is proved for a large class of theories in \cite{Papadimitriou:2004ap}.)

\section{Interacting scalar field}
\label{sec:nologs}
In this section we will be concerned with the holographic renormalization for an interacting scalar field $\Phi$ with the action:
\be
S = \int d^{d+1} x \sqrt{G} \Big( \hf \del_\m \Phi \del^\m \Phi + \hf m^2 \Phi^2 + \frac{1}{3} \lambda \Phi^3\Big)\,.
\ee
We take the background spacetime to be empty Euclidean AdS with a metric of the form:
\be
\label{eq:bgmetric}
G_{\m \n}dx^\m dx^\n = dr^2 + \g_{ij} dx^i dx^j \qquad \qquad \g_{ij} = e^{2r} \d_{ij}\,,
\ee
with $i \in \{1, \ldots, d\}$. The equation of motion becomes:
\be
\label{eq:eomphi}
\square_G \Phi - m^2 \Phi - \lambda \Phi^2 = \ddot \Phi + (\Dp + \Dm) \dot \Phi + e^{-2r} \square_0 \Phi + \Dp \Dm \Phi - \lambda \Phi^2 = 0\,,
\ee
where a dot denotes a radial derivative and we introduced $\Dpm = \hf ( d \pm \sqrt{ d^2 + 4 m^2})$ and $\square_0 = \delta^{ij} \del_i \del_j$. We will below also use the covariant Laplacian $\square_\g = \g^{ij} \del_i \del_j = e^{-2r} \d^{ij} \del_i \del_j$. Notice that $\Dp = \D$ is the scaling dimension of the dual operator and $\Dp + \Dm = d$. We will solve equation \eqref{eq:eomphi} only perturbatively in $\lambda$, ignoring the backreaction onto the metric or any other fields as well as any non-perturbative effects. To this end we expand the solution $\Phi$ as:
\be
\Phi = \Phi\as{0} + \Phi\as{1} + \Phi\as{2} + \Phi\as{3} + \ldots
\ee   
with the individual terms given by the solutions to linear equations:  
\be
\label{eq:eomseries}
\begin{split}
&(\square_G - m^2) \Phi\as{0} = 0\\
&(\square_G - m^2) \Phi\as{1} = \lambda \Phi^2\as{0}\\
&(\square_G - m^2) \Phi\as{2} = 2 \lambda \Phi\as{0} \Phi\as{1}\\
&(\square_G - m^2) \Phi\as{3} =  \lambda ( \Phi\as{1}^2 + 2 \Phi\as{0} \Phi\as{2})\\
&\ldots
\end{split}
\ee
with boundary conditions which can schematically be written as:
\be
\begin{split}
&\Phi\as{0} = \phi\ab{0}{\Dm} e^{-\Dm r} + \ldots\\
&\Phi\as{1} = 0 \, e^{-\Dm r} + \ldots\\
&\Phi\as{2} = 0 \, e^{-\Dm r} + \ldots\\
&\ldots
\end{split}
\ee
We will make these boundary conditions more precise below.

The on-shell action takes the form:
\be
\label{eq:sonshell}
S = - \frac{\lambda}{6} \int d^{d+1}x \sqrt{G} \Phi^3 + \hf \int d^d x \sqrt{\g} \dot \Phi \Phi \,,
\ee
where for our choice of background coordinate system we find that $\sqrt{G} = \sqrt{\g} = e^{dr}$.

\subsection{Free-field solution and two-point function}
In this section we will perform the holographic analysis to zeroth order in $\lambda$. The scalar field is then free and the analysis is familiar. We will here briefly review this analysis and we refer to the lecture notes \cite{Skenderis:2002wp} for more details. Notice that we will not use the more streamlined framework of Hamiltonian holographic renormalization of \cite{Papadimitriou:2004ap} (see also \cite{deBoer:1999xf,deBoer:2000cz,Martelli:2002sp} for related work) as we found it to obfuscate several important features of the analysis.

\subsubsection{Structure of the asymptotic solution}
The solution to the free-field equation, so the first equation in \eqref{eq:eomseries}, is well-known. Asymptotically it takes the form:
\be
\label{eq:phias0}
\Phi\as{0} =  e^{-\Dm r} (\phi\ab{0}{\Dm}+ \phi\ab{0}{\Dm + 2} e^{-2r} + \ldots ) + e^{-\Dp r} (\phi\ab{0}{\Dp} + \phi\ab{0}{\Dp + 2} e^{-2r} + \ldots )\,.
\ee
A subscript $\{k\}$ indicates the order in $\lambda$ at which we are working, whereas a subscript $(k)$ indicates the power of $\exp(-r)$ in the radial expansion.\footnote{Our subscripts in parentheses are deliberately labeled differently from the usual conventions employed in the literature. For example, the leading term $\phi\ab{0}{\Dm}$ would usually have been labeled $\phi\ab{0}{0}$. This kind of labeling however turns out to be rather inconvenient below. For the same reason we introduced $\Dp$ and $\Dm$ rather than using $\D = \Dp$ and $d- \D = \Dm$.} From the free equation of motion one may recursively determine that for $k \in \{1,2,\ldots\}$:
\be
\label{eq:phiab0}
\phi\ab{0}{\Dpm + 2k} = - f(\Dpm + 2k) \square_0 \phi\ab{0}{\Dpm + 2k - 2}\,,
\ee
with:
\be
\label{eq:f}
f(\a) = \frac{1}{(\a - \Dp)(\a - \Dm)} \,.
\ee
This expression is no longer valid at the poles of $f(\Dpm + 2k)$, which occur when $\D = \frac{d}{2} + k$. In those cases we would need a logarithmic term in the radial expansion and we have to deal with these cases separately. We will consider such cases in a separate paper \cite{me}.

According to the standard AdS/CFT dictionary, the component $\phi\ab{0}{\Dm}$ is regarded as the source for the dual operator. On the other hand, $\phi\ab{0}{\Dp}$ is expressable in terms of $\phi\ab{0}{\Dm}$ only by demanding regularity in the interior and will eventually define the one-point function of the dual operator.

\subsubsection{Renormalization of the on-shell action}
To zeroth order in $\lambda$, the bare on-shell action \eqref{eq:sonshell} takes the form:
\be
S\as{0} = \hf \int d^d x \sqrt{\g} \dot \Phi\as{0} \Phi\as{0}\,,
\ee
where $\sqrt{\g} = \exp(dr) = \exp((\Dp + \Dm) r)$ and the integral is taken on a fixed cutoff surface of constant large $r$. Upon subtitution of \eqref{eq:phias0} we find an asymptotic expansion of the form:
\be
S\as{0} = - \hf \int d^d x \,e^{(\Dp  - \Dm) r} ( \Dm \phi\ab{0}{\Dm}^2  +  (2 \Dm + 2) e^{-2r} \phi\ab{0}{\Dm} \phi\ab{2}{\Dm} + \ldots )\,.
\ee
To find the counterterms we follow the usual methods of holographic renormalization and express the divergences in terms of the field $\Phi$ rather than the source $\phi\ab{0}{\Dm}$. In order to do so we should write $\phi\ab{0}{\Dm}$ as a function of $\Phi$. To leading order this relation simply reads:
\be
\label{eq:phi0Phi}
\phi\ab{0}{\Dm} = e^{\Dm r} \Phi + \ldots
\ee
We then substitute this inverted series into the divergent part of the action. The leading terms can then be simply removed by adding a counterterm action of the same form, see \cite{Skenderis:2002wp} for details. This procedure leads to a counterterm action
\be
\label{eq:Sct0}
S\asct{0} = \int d^d x \sqrt{\g} (\hf \Dm \Phi^2 - f(\Dm + 2) \Phi \square_\g \Phi + \ldots)\,,
\ee
where the dots represent terms of the form $c_k \Phi \square_\g^k \Phi$ for $1 < k < (\Dp - \Dm)/2$. The coefficients $c_k$ can be determined recursively but their exact value will not be needed here. After adding the counterterms the total action $S + S\asct{0}$ is finite as we send the cutoff $r \to \infty$.

\subsubsection{One-point function}
To find the one-point function we look at the first variation of the total action,
\be
\delta (S + S\asct{0}) = \int d^d x \sqrt{\g} \Pi_r \delta \Phi \,.
\ee
where we introduced:
\be
\label{eq:pir}
\Pi_r = \dot \Phi + \Dm \Phi - 2 f(2\Dm + 2) \square_\g \Phi + \ldots\,,
\ee
which is called the \emph{renormalized conjugate momentum} in \cite{Papadimitriou:2004ap}. The more subleading terms in $\Pi_r$ are by construction obtained by taking the variation of the counterterms but we will again not need their exact expressions here. We find that the second and higher terms in \eqref{eq:pir} precisely strip off all the locally determined terms in $\dot \Phi$ so that the leading term in the radial expansion is
\be
\label{eq:parenthesis0}
\Pi_r = (\Dm - \Dp) e^{-\Dp r} \phi\ab{0}{\Dp} + \ldots 
\ee
Furthermore,
\be
\label{eq:deltaphi0}
\delta \Phi = e^{-\Dm r} \delta \phi\ab{0}{\Dm} + \ldots
\ee
so we obtain, up to terms which vanish as $r \to \infty$, that
\be
\delta (S + S\asct{0}) = \int d^d x (\Dm - \Dp) \phi\ab{0}{\Dp} \delta \phi\ab{0}{\Dm}\,,
\ee
where we used that $\Dm + \Dm = d$. The one-point function in the presence of sources is therefore given by:
\be
\label{eq:vev2pt}
\vev{\op} = \lim_{r \to \infty} e^{\Dp r} \Pi_r =  (\Dm - \Dp) \phi\ab{0}{\Dp}\,,
\ee
which also agrees with \cite{Skenderis:2002wp}.

\subsection{First-order correction}
We will now extend our analysis to first order in $\lambda$. The procedure here is exactly the same as in the previous section: we find the asymptotic form of the solution, substitute it into the on-shell action, compute the counterterm action which cancels the divergences and finally compute the first-order variation of the renormalized on-shell action to find the finite one-point function in the presence of sources.

\subsubsection{Structure of the asymptotic solution}
From the second line in \eqref{eq:eomseries} we find a radial expansion for $\Phi\as{1}$ of the form:
\be
\label{eq:expphi1}
\begin{split}
\Phi\as{1} &= e^{-2 \Dm r} ( \phi\ab{1}{2\Dm} + e^{-2r} \phi\ab{1}{2\Dm + 2} + \ldots)\\
&\qquad + e^{- \Dm r} ( \phi\ab{1}{\Dm} + \ldots )\\
&\qquad + e^{- (\Dm + \Dp)r} (\phi\ab{1}{\Dm + \Dp} + \ldots)\\
&\qquad + e^{- \Dp r} (\phi\ab{1}{\Dp} + \ldots)\\
&\qquad + e^{-2 \Dp r}( \phi\ab{1}{2\Dp} + \ldots)\,,
\end{split}
\ee
with coefficients given by: 
\be
\begin{split}
\phi\ab{1}{2 \Dm} &= f(2\Dm) \lambda \phi\ab{0}{\Dm}^2\\
\phi\ab{1}{2\Dm + 2} &= f(2\Dm + 2) (2 \lambda \phi\ab{0}{\Dm} \phi\ab{0}{\Dm + 2} - \square_0 \phi\ab{1}{2\Dm})\\
\phi\ab{1}{2\Dm + 4} &= f(2\Dm + 4) (2 \lambda \phi\ab{0}{\Dm} \phi\ab{0}{\Dm + 4} + \lambda \phi\ab{0}{\Dm + 2}^2 - \square_0 \phi\ab{1}{2\Dm + 2})\\
\end{split}
\ee
and similarly for $\Dm \to \Dp$. We also find that
\be
\phi\ab{1}{\Dm + \Dp} = 2 f(\Dm + \Dp) \lambda \phi\ab{0}{\Dm} \phi\ab{0}{\Dp}\,.
\ee
Notice that $f(2\Dm + 2k)$ has a pole if $\D = 2 k + d$ for $k \in \{1,2,3,\ldots\}$. This again signifies that new logarithmic terms need to be added to the radial expansion and we will explore these cases in \cite{me}. In this paper we will always assume that $\D$ is such that no logarithmic terms need to be added to the radial expansion, also when working at higher orders in $\lambda$ below. 

The above procedure recursively determines all the terms in \eqref{eq:expphi1} except for the terms $\phi\ab{1}{\Dpm}$. This is as expected: the coefficient $\phi\ab{1}{\Dm}$ represents a change of order $\lambda$ to the source term $\phi\ab{0}{\Dm}$. Since we want to keep the sources fixed, we require it to be zero:
\be
\phi\ab{1}{\Dm} = 0\,.
\ee
This is the more precise boundary condition for $\Phi\as{1}$ which we referred to above. (Similarly, at higher orders the precise boundary condition for $\Phi\as{k}$ with $k > 1$ is that $\phi\ab{k}{\Dm} = 0$.) Last, the term $\phi\ab{1}{\Dp}$ represents a change to the vev term $\phi\ab{0}{\Dp}$. This term can only be obtained from the complete solution at order $\lambda$ to the equation of motion and just like $\phi\ab{0}{\Dp}$ will only follow from demanding regularity in the interior.

Let us remark that the asymptotics \eqref{eq:expphi1} are not easily obtained from the possibly more familiar analysis which uses a bulk-bulk propagator to construct $\Phi\as{1}$. In appendix \ref{app:bulkbulk} we explain in more detail how these terms can nevertheless be recovered by using an appropriate Fourier-transformed expression for the bulk-bulk propagator.

\subsubsection{Renormalization of the on-shell action}
Let us now substitute the solution $\Phi\as{0} + \Phi\as{1}$ into the on-shell action. To first order in $\lambda$ we find that \eqref{eq:sonshell} becomes:
\be
S\as{1} = - \frac{\lambda}{6} \int d^{d+1}x \sqrt{G} \Phi\as{0}^3 + \hf \int d^d x \sqrt{\g} (\dot \Phi\as{0} \Phi\as{1} + \dot \Phi\as{1}\Phi\as{0})\,.
\ee
To cancel the divergences at zeroth order we added above the counterterm action $S\asct{0}$ given in \eqref{eq:phias0}. Since this counterterm action is a function of $\Phi$ and not just of $\Phi\as{0}$, it also has an expansion in $\lambda$ which to first order takes the form:
\be
S\asctexp{0}{1} =\int d^d x \sqrt{\g} (\Dm \Phi\as{0}\Phi\as{1} + 2f(\Dm + 2) \Phi\as{0} \square_\g \Phi\as{1} + \ldots)\,.
\ee
To re-emphasize, the subscript on the left-hand side symbolizes the expansion to first order in $\lambda$ of the counterterm action $S\asct{0}$ which rendered the on-shell action finite to zeroth order in $\lambda$. Substitution of the above radial expansions then results in the following divergent terms:
\be
\label{eq:S1div}
\begin{split}
S\as{1} + S\asctexp{0}{1} = \lambda \int d^d x \Big( &e^{(\Dp - 2\Dm)r} \frac{1}{3(\Dp - 2 \Dm)}\phi\ab{0}{\Dm}^3 + \ldots \\
& + e^{-\Dm r} \frac{1}{\Dp - 2\Dm} \phi\ab{0}{\Dm}^2 \phi\ab{0}{\Dp} + \ldots \Big)\,.
\end{split}
\ee

In this expression the dots represent terms involving powers of the box $\square$. Such terms are actually rather unimportant for our analysis. This is because they sit at powers of $\exp(- n \Dm - m \Dp - 2 k r)$ for nonzero positive $k$ and (provided there are no logarithmic terms) these powers do not mix with the terms with $k = 0$. This shows that the terms involving boxes cannot contribute, for example, to the one-point function, and can also be more or less ignored for the discussion below. We will therefore often not write such terms explicitly and rather use dots to indicate that we omitted them.

A remarkable property of \eqref{eq:S1div} is that the nonlocally determined part $\phi\ab{0}{\Dp}$ appears explicitly in the divergent part of the action. This may seem worrying, since non-local divergences may require non-local counterterms and such counterterms generally spoil the predictability of the theory. Although we may expect such non-local divergences for finite values of the source $\phi\ab{0}{\Dm}$, we certainly do not expect them while treating the sources infinitesimally as we do here.

However the AdS/CFT dictionary is slightly more subtle than the above reasoning suggests and the fact that the divergences are non-local when expressed in terms of $\phi\ab{0}{\Dm}$ is in itself not problematic. This is because the proper boundary data for a finite cutoff is the field $\Phi$ and not $\phi\ab{0}{\Dm}$. To zeroth order in $\lambda$, the relation between $\Phi$ and $\phi\ab{0}{\Dm}$ is locally determined to order $\exp(-\Dp r)$ in the radial expansion. However to first order in $\lambda$ this local relation is spoiled already at order $\exp(-(\Dp + \Dm) r)$ by the appearance of the nonlocal term $\phi\ab{1}{\Dp + \Dm}$ in \eqref{eq:expphi1}. As long as $\Dm < 0$, so as long as the operator is irrelevant, this new nonlocal term is more leading than the old one. It is then not surprising that (at order $\lambda$) a divergence which may be local in $\Phi$ appears non-local when expressed in terms of $\phi\ab{0}{\Dm}$. The proper method is therefore to express the divergences first in terms of $\Phi$ and only then assess whether they are local or not.

For the case at hand we find that the non-local divergence is actually cancelled by the leading counterterm. This occurs as follows. We first use the relation \eqref{eq:phi0Phi} between the source $\phi\ab{0}{\Dm}$ which we recall takes the form:
\be
\phi\ab{0}{\Dm} = e^{\Dm r} \Phi + \ldots\,,
\ee
where the dots involve terms with boxes as well as higher order terms in $\lambda$ which are all unimportant to us here. The leading divergence in \eqref{eq:S1div} therefore requires a counterterm of the form:
\be
S\asct{1} = - \lambda \int d^d x \sqrt{\g} \Big( \frac{1}{3(\Dp - 2 \Dm)} \Phi^3 + \ldots \Big) \,,
\ee
where the dots represent the terms involving boxes which cancel similar divergences in \eqref{eq:S1div}. Now, upon substitution of the radial expansion for $\Phi$ (and keeping only the terms to first order in $\lambda$), we find that this counterterm precisely cancels \emph{also} the `non-local' divergence on the second line of \eqref{eq:S1div}. Therefore, the combined divergence in \eqref{eq:S1div} was in fact local when expressed in terms of $\Phi$ and therefore not problematic.

We have verified using Mathematica that a similar cancellation occurs for many subleading `pseudo-non-local' divergences involving up to four boxes which we represented by the dots in equation \eqref{eq:S1div}. For example, the second counterterm in $S\asct{1}$, which actually takes the form:
\be
\lambda \int d^d x \sqrt{\g} \frac{\Phi^2 \square_\g \Phi}{(2 + \Dm -\Dp)(2\Dm - \Dp)(2 + 2 \Dm - \Dp)}
\ee
cancels both the local term of order $\exp(-(2\Dm - \Dp + 2)r)$ and the pseudo-non-local term of order $\exp(-(\Dm +2)r)$ in the radial expansion of $S\as{1} + S\asctexp{0}{1}$. To first order in $\lambda$, then, it is natural to claim that all divergences can be cancelled with counterterms which are local in $\Phi$. This would be in agreement with the locality of the divergences on the field theory side, but unfortunately we were unable to completely prove this result. A full demonstration can presumably be obtained by a suitable modification of the radial Hamiltonian approach to holographic renormalization which was worked out in detail in \cite{Papadimitriou:2004ap}.

As a sidenote, let us remark that one crucial feature of this radial Hamiltonian approach is to replace the radial derivative by a covariant functional differentiation operator which approximates the radial derivative near the boundary. In our case, however, the asymptotic behavior of $\Phi$ is modified to each order in $\lambda$ and this requires corresponding perturbative adjustments to any functional differentiation operator which asymptotes to the radial derivative. This considerably complicates the usual analysis, especially when one attempts to demonstrate the general statements of the previous paragraph.

\subsubsection{One-point function}
The first-order variation of the renormalized on-shell action now takes the form:
\be
\label{eq:deltaS1}
\delta(S + S\asct{0} + S\asct{1}) = \int d^d x \sqrt{\g} (\dot \Phi + \Dm \Phi - \frac{\lambda}{\Dp - 2 \Dm} \Phi^2 + \ldots) \delta \Phi\,.
\ee
The term in parentheses is an extension of the renormalized conjugate momentum $\Pi_r$ defined in \eqref{eq:pir} to include the variation of the counterterms at first order in $\lambda$. In this case the leading term in its radial expansion is however \emph{not} directly related to the one-point function. Indeed, we find that
\be
\label{eq:pi1}
\begin{split}
& \dot \Phi + \Dm \Phi - \frac{\lambda}{\Dp - 2 \Dm} \Phi^2 + \ldots = 
(\Dm - \Dp) (\phi\ab{0}{\Dp} + \phi\ab{1}{\Dp} ) e^{-\Dp r} \\& \qquad  - 2 \lambda f(2\Dm) (\Dm - \Dp) \phi\ab{0}{\Dm} \phi\ab{0}{\Dp} e^{-(\Dm+ \Dp) r} + \ldots \,,
\end{split}
\ee
which has an unwanted divergent term of order $\exp(-(\Dp + \Dm)r)$. However in \eqref{eq:deltaS1} this term gets multiplied by $\delta \Phi$, which is given by
\be
\label{eq:deltaphi1}
\delta \Phi = ( 2 \lambda f(2\Dm) \phi\ab{0}{\Dm} e^{-\Dm r} + 1  + \ldots) \delta \phi\ab{0}{\Dm} e^{-\Dm r}\,,
\ee
as follows directly from \eqref{eq:expphi1}. Combining the above two equations we eventually find that the unwanted term precisely cancels and we obtain the finite first-order variation:
\be
\label{eq:deltaStot1}
\delta(S + S\asct{0} + S\asct{1}) = \int d^d x (\Dm - \Dp) (\phi\ab{0}{\Dp} + \phi\ab{1}{\Dp}) \d \phi\ab{0}{\Dm}\,.
\ee
The one-point function is therefore again just $(\Dm - \Dp)$ times the normalizable mode,
\be
\label{eq:vev3pt}
\vev{\op} = (\Dm - \Dp) (\phi\ab{0}{\Dp} + \phi\ab{1}{\Dp})\,,
\ee
which is the familiar result. (We refer to \cite{Skenderis:2002wp} for an explanation of the fact that \eqref{eq:vev3pt}, or more precisely the all-orders extension $\vev{\op} = (\Dm - \Dp) \phi\bs{\Dp}$, leads to the usual Witten diagram expression for three- and higher-point functions.)

Notice that in almost all currently known examples of holography the relation between the source and the bulk field always takes a simple form like \eqref{eq:deltaphi0}, so the source $\phi\ab{0}{\Dm}$ is always the leading term in the radial expansion of the field $\Phi$. For irrelevant operators this no longer has to be the case since we see explicitly from \eqref{eq:deltaphi1} that the relation between the field and the source becomes more involved. Comparing \eqref{eq:pi1} with \eqref{eq:parenthesis0} we find that this also has repercussions on the relation between the renormalized conjugate momentum and the one-point function which is given by $\phi\ab{0}{\Dp}$.

\subsection{Second-order correction}
At second order in $\lambda$ we should formally start taking into account the backreaction onto the metric of the scalar field, at least within any string theory context where all coupling constants are (generically) of the same order. However we expect that a more complete analysis which includes the other supergravity fields will not significantly alter the procedure of holographic renormalization. As we announced in the introduction, we will therefore continue to set to zero all the other fields. Within this setting the second-order correction in $\lambda$ follows largely the analysis of the first-order correction. We will therefore be rather brief in this subsection and omit most of the details.

First of all, from the third line in \eqref{eq:eomseries}, we find the asymptotic solution to the equation of motion at second order in $\lambda$:
\be
\label{eq:expphi2}
\begin{split}
\Phi\as{2} &= e^{-3 \Dm r}(\phi\ab{2}{3\Dm} + e^{-2r} \phi\ab{2}{3\Dm + 2} + \ldots)\\
&\qquad + e^{-(2\Dm + \Dp)r}(\phi\ab{2}{2\Dm + \Dp} + \ldots)\\
&\qquad + e^{-\Dp r}(\phi\ab{2}{\Dp} + \ldots)\\
&\qquad + e^{-(\Dm + 2\Dp)r}(\phi\ab{2}{\Dm + 2\Dp} + \ldots)\\
&\qquad + e^{- 2 \Dp r}(\phi\ab{2}{2\Dp}+ \ldots)\\
&\qquad + e^{- 3 \Dp r}(\phi\ab{2}{3\Dp} + \ldots)\,,
\end{split}
\ee
where we ordered the terms from larger to smaller powers of $e^r$ for $\Dm < 0$ and the dots represent subleading terms with boxes. We again set to zero a possible higher-order correction to the source, which would be a term $\exp(-\Dm r) \phi\ab{2}{\Dm}$. In this expansion, all the coefficients except for $\phi\ab{2}{\Dp}$ can be (recursively) determined by an asymptotic analysis in terms of those entering in $\Phi\as{0}$ and $\Phi\as{1}$, and therefore eventually in terms of $\phi\ab{0}{\Dpm}$ and $\phi\ab{1}{\Dp}$. For example, we find for the leading term:
\be
\label{eq:phi23dm}
\begin{split}
\phi\ab{2}{3\Dm} = 2 \lambda^2 f(3\Dm) f(2\Dm) \phi\ab{0}{\Dm}^3\,.
\end{split}
\ee
The exceptional term $\phi\ab{2}{\Dp}$ of course will eventually become the higher-order correction to the one-point function.

The divergences in the on-shell action now take the form:
\be
\label{eq:Sexp2}
\begin{split}
S\as{2} + S\asctexp{0}{2} + S\asctexp{1}{2} = \lambda^2 \int d^d x \Big( &e^{(\Dp - 3 \Dm) r}
\frac{ \phi\ab{0}{\Dm}^4 }{2(3\Dm - \Dp)(2 \Dm - \Dp)^2} + \ldots \\
&+ e^{ - 2 \Dm r} \frac{2 \phi\ab{0}{\Dm}^3 \phi\ab{0}{\Dp} }{(3\Dm - \Dp)(2 \Dm - \Dp)^2}   + \ldots\Big)\,,
\end{split} 
\ee
where the dots again represent terms involving powers of boxes. Notice that the structure of the divergence is very similar to \eqref{eq:S1div}. The next possible subleading term without boxes is proportional to $\phi\ab{0}{\Dm}^2 \phi\ab{0}{\Dp}^2 \exp(-(\Dm + \Dp) r)$ and therefore convergent as $r \to \infty$.


To find the first counterterm it suffices again to use the leading-order inversion \eqref{eq:phi0Phi} and the counterterm which cancels the leading divergence is then found to be:
\be
S\asct{2} = - \lambda^2 \int d^d x \sqrt{\g} \Big(
\frac{\Phi^4}{2(3\Dm - \Dp)(2 \Dm - \Dp)^2}  + \ldots\Big)\,.
\ee
Again, this counterterm also cancels the pseudo-non-local term on the second line of \eqref{eq:Sexp2} and we have checked with Mathematica that a similar phenomenon occurs for the subleading counterterms involving up to two boxes. We again suppose that this cancellation is systematic, thus extending our claim of the previous section to second order in $\lambda$. (Notice that in the next subsection we will conclude that not all divergences involving $\phi\ab{0}{\Dp}$ are pseudo-non-local.)

The one-point function is obtained by expanding $\delta S$ to second order in $\lambda$. The relation between the source $\phi\ab{0}{\Dm}$ and the field $\Phi$ is now given by:
\be
\label{eq:deltaphi2}
\begin{split}
\delta \Phi &= \Big( 6  \lambda^2 f(3\Dm) f(2\Dm) \phi\ab{0}{\Dm}^2 e^{-2\Dm r} \\ & \qquad \qquad+ 2 \lambda f(2\Dm) \phi\ab{0}{\Dm} e^{-\Dm r} + 1  + \ldots\Big) \delta \phi\ab{0}{\Dm} e^{-\Dm r}\,,
\end{split}
\ee
which follows from \eqref{eq:deltaphi1}, \eqref{eq:expphi2} and \eqref{eq:phi23dm}. Just as in equation \eqref{eq:pi1}, we find similar terms in the expression for the extension of the renormalized conjugate momentum to second order in $\lambda$ which precisely cancel the two leading terms in \eqref{eq:deltaphi2}. The total variation of the action to second order in $\lambda$ then again takes the simple and finite form:
\be
\delta(S + S\asct{0} + S\asct{1} + S\asct{2}) = \int d^d x (\Dm - \Dp) (\phi\ab{0}{\Dp} + \phi\ab{1}{\Dp} + \phi\ab{2}{\Dp}) \d \phi\ab{0}{\Dm}
\ee
and the renormalized one-point function up to this order finally becomes
\be
\vev{\op} = (\Dm - \Dp) (\phi\ab{0}{\Dp} + \phi\ab{1}{\Dp} + \phi\ab{2}{\Dp})\,,
\ee
again as expected.

\subsection{Third-order correction}
We will now consider the third-order correction where we will see that extra complications arise and the variational principle has to be modified to include also the variation of the conjugate momentum $\dot \Phi$ at the boundary.

\subsubsection{Structure of the asymptotic solution}
The third-order correction to the solution of the equation of motion takes a rather complicated form:
\be
\begin{split}
\Phi\as{3} &= e^{-4 \Dm r}(\phi\ab{3}{4\Dm} + e^{-2r} \phi\ab{3}{4\Dm + 2} + \ldots)\\
&\qquad + e^{-(3\Dm + \Dp)r}(\phi\ab{3}{3\Dm + \Dp} + \ldots)\\
&\qquad + e^{-(2\Dm + \Dp)r}(\phi\ab{3}{2\Dm + \Dp} + \ldots)\\
&\qquad + e^{-(2\Dm + 2\Dp)r}(\phi\ab{3}{2\Dm + 2\Dp} + \ldots)\\
&\qquad + e^{-(\Dm + 2 \Dp)r}(\phi\ab{3}{\Dm + 2 \Dp} + \ldots)\\
&\qquad + e^{-(\Dm + 3 \Dp)r}(\phi\ab{3}{\Dm + 3 \Dp} + \ldots)\\
&\qquad + e^{-\Dp r}(\phi\ab{3}{\Dp} + \ldots)\\
&\qquad + e^{- 2 \Dp r}(\phi\ab{3}{2\Dp}+ \ldots)\\
&\qquad + e^{- 3 \Dp r}(\phi\ab{3}{3\Dp} + \ldots)\\
&\qquad + e^{- 4 \Dp r}(\phi\ab{3}{4\Dp} + \ldots)\,.
\end{split}
\ee
Notice that we again set to zero a possible higher-order correction to the source, which would be a term $\exp(-\Dm r) \phi\ab{3}{\Dm}$. In this expansion, all the coefficients except for $\phi\ab{3}{\Dp}$ can be determined by the asymptotic analysis in terms of those entering in $\Phi\as{0}$, $\Phi\as{1}$ and $\Phi\as{2}$ and so eventually in terms of $\phi\ab{0}{\Dpm}$, $\phi\ab{1}{\Dp}$ and $\phi\ab{2}{\Dp}$. The exceptional non-locally determined term $\phi\ab{3}{\Dp}$ will again eventually become the higher-order correction to the one-point function.

We find for the leading coefficient:
\be
\phi\ab{3}{4\Dm} = \lambda^3 f(4\Dm) [f(2\Dm)^2 + 4 f(3\Dm) f(2\Dm)] \phi\ab{0}{\Dm}^4 
\ee
and for the first subleading term not involving boxes:
\be
\phi\ab{3}{3\Dm + \Dp} = \lambda^3 c_{3,1} \phi\ab{0}{\Dm}^3 \phi\ab{0}{\Dp}\,,
\ee
with
\be
\label{eq:c31}
\begin{split}
c_{3,1} &= 
4  f(3\Dm + \Dp) \Big( 2 f(2\Dm + \Dp) f(\Dm + \Dp) + f(2\Dm)[f(3\Dm) \\&\qquad  + f(2\Dm + \Dp) + f(\Dm + \Dp)]\Big) \\
&= 
\frac{4 \left(9 \Dm -\Dp \right)}{3 \Dm ^2 \Dp  \left(12 \Dm ^4+8 \Dm ^3 \Dp -7 \Dm ^2 \Dp ^2-2 \Dm  \Dp ^3+\Dp ^4\right)}\,.
\end{split}
\ee
Notice that $\phi\ab{3}{3\Dm + \Dp}$ involves the nonlocally determined term $\phi\ab{0}{\Dp}$ and for $3\Dm + \Dp < \Dm$, so for $\D > 2d$, this term is \emph{more} leading than the source term $\phi\ab{0}{\Dm}$. Therefore, the asymptotic relation between the field $\Phi$ and the source term $\phi\ab{0}{\Dm}$ will at this order involve $\phi\ab{0}{\Dp}$ and therefore become non-local. As we shall see below, the holographic renormalization procedure will resolve this issue by an appropriate modification of the variational principle which will ensure precisely that $\phi\ab{0}{\Dm}$ remains the correct source term also at this order.

\subsubsection{Renormalization of the on-shell action}
\label{sec:vev5pt}
In the on-shell action we now find the following divergent terms:
\begin{multline}
\label{eq:S3div}
 S\as{3} + S\asctexp{0}{3} + S\asctexp{1}{3} + S\asctexp{2}{3} = \\  - 2 \lambda^3 \int d^d x \Big( \frac{1}{\left(2 \Dm -\Dp \right){}^3 \left(12 \Dm^2-7 \Dm \Dp +\Dp^2\right)} \phi\ab{0}{\Dm}^5 e^{(-4\Dm + \Dp)r}\\
  +  \frac{5}{\left(2 \Dm -\Dp \right){}^3 \left(12 \Dm^2-7 \Dm \Dp +\Dp^2\right)} \phi\ab{0}{\Dm}^4 \phi\ab{0}{\Dp} e^{(-3\Dm + \Dp)r}\\
- \frac{\left(10 \Dm^3-63 \Dm^2 \Dp+26 \Dm \Dp^2-3 \Dp^3\right)  }{3 \Dm^3 \left(2 \Dm-\Dp\right){}^3 \left(6 \Dm^2+\Dm \Dp-\Dp^2\right)}  \phi\ab{0}{\Dm}^3 \phi\ab{0}{\Dp}^2 e^{-(2\Dm + \Dp)r}
+ \ldots \Big)\,.
\end{multline}
We again omitted the terms involving boxes. Notice that the last term is only divergent when $2 \Dm +  \Dp < 0$, that is when:
\be
\label{eq:conddelta}
\D > 2 d\,,
\ee
which we will henceforth assume to be the case.

As before, to cancel the leading divergence we need the relation \eqref{eq:phi0Phi} which says that to leading order in both $\lambda$ and the radial expansion:
\be
\phi\ab{0}{\Dm} = e^{\Dm r}\Phi + \ldots 
\ee
The counterterm cancelling the leading divergence is therefore:
\be
S\asct{3} = 2 \lambda^3 \int d^d x \sqrt{\g} \frac{\Phi^5}{\left(2 \Dm -\Dp \right){}^3 \left(12 \Dm^2-7 \Dm \Dp +\Dp^2\right)} + \ldots
\ee
and this counterterm again also happens to cancel the divergence on the second line of \eqref{eq:S3div} which was therefore only pseudo-non-local. However after adding this counterterm we find that the other non-local divergence remains:
\be
\label{eq:S3stilldiv}
\begin{split}
&S\as{3} + S\asctexp{0}{3} + S\asctexp{1}{3} + S\asctexp{2}{3} + S\asctexp{3}{3} =\\ 
& \qquad \lambda ^3 \int d^d x \, c_{3,2} \phi\ab{0}{\Dm}^3 \phi\ab{0}{\Dp}^2 e^{- (2\Dm + \Dp)r} + \ldots
\end{split}
\ee
whose coefficient is now modified to:
\be
\begin{split}
c_{3,2} = \frac{2  \left(\Dm-\Dp\right){}^2 \left(100 \Dm^2-32 \Dm \Dp+3 \Dp^2\right)}{3 \Dm^3 \left(2 \Dm-\Dp\right){}^3 \left(24 \Dm^3-2 \Dm^2 \Dp-5 \Dm \Dp^2+\Dp^3\right)}\,.
\end{split}
\ee
Similar subleading divergences remain if we add the counterterms involving boxes to cancel divergences of the symbolic form $\square^k \phi\ab{0}{\Dm}^5$: although the counterterms with boxes again happen to cancel divergences of the form $\square^k \phi\ab{0}{\Dm}^4\phi\ab{0}{\Dp}$, just as they did for lower orders in $\lambda$, they do not cancel the possibly divergent terms of the form $\square^k \phi\ab{0}{\Dm}^3 \phi\ab{0}{\Dp}^2$. As we discussed when we computed the first-order correction, the counterterms involving boxes will also never mix with or change the divergence \eqref{eq:S3stilldiv}.

It follows that we need a new counterterm to cancel the divergence in \eqref{eq:S3stilldiv}. As always, in order to find it we have to write the divergence in terms of the induced fields by inverting the asymptotic relation between the fields and the boundary sources. Since the divergence already has an overall factor $\lambda^3$, it suffices to perform this inversion to order $\lambda^0$. The divergence in \eqref{eq:S3stilldiv} however involves the non-locally determined term $\phi\ab{0}{\Dp}$ and we cannot use \eqref{eq:phi0Phi} as we did for all the counterterms so far. Rather, to express $\phi\ab{0}{\Dp}$ in terms of the boundary fields we need the radial momentum $\dot \Phi$. Indeed, the only possibility is to use:
\be
\phi\ab{0}{\Dp} = \frac{e^{\Dp r}}{\Dm - \Dp} (\dot \Phi + \Dm \Phi + \ldots) \equiv \frac{e^{\Dp r}}{\Dm - \Dp} \Pi_r + \ldots\,,
\ee
where the dots represent terms of the form $\square_\g^k \Phi$ which subtract the terms of the form $\square_0^k \phi\ab{0}{\Dm}^k$ from the expansion \eqref{eq:phias0}. The expression in parentheses is precisely the renormalized conjugate momentum defined in \eqref{eq:pir}. From its definition we see that it is a covariant function of the field $\Phi$ and the bare conjugate momentum $\dot \Phi$. Using $\Pi_r$ we directly find that the extra counterterm has the form: 
\be
\label{eq:tildeSct3}
\tilde S\asct{3} = - \frac{\lambda^3}{(\Dm - \Dp)^2} \int d^d x \sqrt{\g} ( c_{3,2} \Phi^3 \Pi_r^2 + \ldots)\,,
\ee
where again the dots represent the terms involving boxes. The counterterm \eqref{eq:tildeSct3} is certainly a non-local function of the boundary data $\Phi$, indicative of a more fundamental phenomenon than the pseudo-non-local terms we found at first and second order.

Despite its non-local nature, the appearance of \eqref{eq:tildeSct3} is completely compatible with the field theory analysis. Namely, by power counting this is precisely the order at which \emph{multi-trace counterterms} should become necessary and according to the standard multi-trace results of \cite{Berkooz:2002ug,Witten:2001ua,Mueck:2002gm,Sever:2002fk,Elitzur:2005kz,Papadimitriou:2007sj} this is implemented holographically by the addition of non-local boundary terms of the form \eqref{eq:tildeSct3} in the bulk theory.

The power counting in the field theory is done as follows. We consider switching on a nonzero source $t(x)$ for a scalar operator $\op$ of the form
\be
\int d^d x \, t(x) \op(x)\,.
\ee
Generally, in order to render the partition function finite one needs to supplement this deformation with all possible counterterms of dimension less than $d$ that are compatible with the remaining symmetries. If the operator $\op$ has dimension $\D$ then $t(x)$ has dimension $d - \D$. The dimension of $t(x)$ is therefore negative for irrelevant operators and we can construct counterterms with operators of arbitrarily high dimension by compensating with a large number of sources: this is the usual phenomenon of non-renormalizibility which forces us to work perturbatively in $t(x)$. At order $t(x)^2$, such counterterms for example take the form:
\be
\Lambda^{\D - 2k - d} \int d^d x \, t(x)^2 \square^k \op
\ee
for all $k < (\D - d)/2$, or a logarithmic counterterm when equality holds. We in general may also find counterterms involving many other operators when operator mixing occurs. In particular, if $\op$ mixes with the identity operator we find additive counterterms of the form:
\be
\Lambda^{2\D - 2k - d} \int d^d x \, t(x) \square^k t(x)\,.
\ee
A logarithmic counterterm of a similar form occurs whenever $\D = k + d/2$ (which is in fact precisely the counterterm that leads to the usual nontrivial conformal anomaly for the two-point function of such operators in a CFT).

Let us now consider the multi-trace operator $\op^2(x)$ and more particularly the counterterm:
\be
\label{eq:ctmultitrace}
\int d^d x \, t^k(x) \op^2(x)\,,
\ee
with $k$ an integer whose lowest possible value is determined as follows. To leading order $\op^2(x)$ has scaling dimension $2\D$. In order to obtain a counterterm of dimension less than $d$, we need $k(d-\D) + 2\D < d$. Assuming that $\D > d$ we find that:
\be
\label{eq:kbound}
k > \frac{2\D - d}{\D - d}\,.
\ee
The right-hand side is a monotonically decreasing function of $\D$ which for large $\D$ tends to $2$ from above. The lowest possible value for the integer $k$ is therefore $k = 3$ and then the integrand in \eqref{eq:ctmultitrace} has dimension less than $d$ for all $\D > 2d$. We thus conclude that this counterterm can only be important when we expand the partition function to at least third order in $t(x)$. However at order $t(x)^3$ this counterterm merely results in the vacuum expectation value of $\op^2(x)$ which vanishes in our background. The contribution at order $t(x)^4$ multiplies the two-point function $\vev{\op^2(x)\op(y)}$ which also vanishes as the operators have different dimensions. The lowest order at which this counterterm may be observed is therefore in the renormalization of the five-point function and this precisely matches the result from the bulk theory.

Indeed, by an analogous power-counting argument in the bulk theory one finds that counterterms involving $\Pi_r$ arise precisely at order $k$ as predicted by \eqref{eq:kbound}. To see this, notice that such a divergence has the form $\lambda^k \phi\ab{0}{\Dm}^k \phi\ab{0}{\Dp}^2$ and therefore has a power $\exp(- [(k-2) \Dm + \Dp]r)$ which becomes positive precisely when \eqref{eq:kbound} is satisfied. Furthermore, in agreement with the above argument it becomes relevant at the level of the ($k+2$)-point function. In particular, by assuming \eqref{eq:conddelta} we obtained such a counterterm in the bulk theory precisely at the level of the five-point function.

The counterterms like \eqref{eq:ctmultitrace}, which we deem as non-local in the gravity theory, are therefore interpreted as \emph{local multi-trace counterterms} in the field theory. In the gravity theory the usual renormalization prescription gets modified: we should not merely look for counterterms which are local expresssions of the boundary data $\Phi$, but upon the appearance of truly non-local divergences we should rather admit the insertion of counterterms involving the (renormalized) conjugate momentum $\Pi_r$ as well. The locality of the field theory counterterms is then more precisely translated into the locality of the counterterm action \emph{as a function of $\Phi$ and $\Pi_r$} rather than as a function of $\Phi$ alone. Notice however that multi-trace counterterms are always at least quadratic in $\Pi_r$ and a term linear in $\Pi_r$ is therefore excluded from appearing in the counterterm action. 
 
The addition of the counterterm \eqref{eq:tildeSct3} changes the variational principle. Correspondingly, we find for the first variation of the on-shell action that:
\be
\label{eq:deltaStot3}
\begin{split}
&\delta ( S + S\asct{0} + S\asct{1} + S\asct{2} + S\asct{3} + \tilde S\asct{3} ) =
\\&\qquad
\int d^d x \sqrt{\g} \Big( \dot \Phi \delta \Phi + \sum_{k = 1}^3 \delta S\asct{k} - \frac{\lambda^3}{(\Dm - \Dp)^2} c_{3,2} ( 3\Phi^2 \Pi_r^2 \delta \Phi + 2 \Phi^3 \Pi_r \delta \Pi_r) + \ldots \Big) \,, 
\end{split}
\ee
with the dots representing the terms involving boxes in $\tilde S\asct{3}$.

As we mentioned below \eqref{eq:c31}, the fact that a change in the variational principle was necessary already demonstrated itself in the relation between $\Phi$ and the source $\phi\ab{0}{\Dm}$. Indeed, to this order in $\lambda$ we find that:
\be
\label{eq:deltaphinonlocal}
\begin{split}
\delta \Phi &= \Big( 16 \lambda^3 f(4\Dm) [f(2\Dm)^2 + 4 f(3\Dm) f(2\Dm)] \phi\ab{0}{\Dm}^3 e^{-3\Dm r}\\
&\qquad + 3 \lambda^3 c_{3,1} \phi\ab{0}{\Dm}^2 \phi\ab{0}{\Dp} e^{-(2\Dm + \Dp)r}
+ 6 \lambda^2 f(3\Dm) f(2\Dm) \phi\ab{0}{\Dm}^2 e^{-2\Dm r} \\&\qquad + 2 f(2\Dm)\lambda \phi\ab{0}{\Dm} e^{-\Dm r} + 1  + \ldots\Big) \delta \phi\ab{0}{\Dm} e^{-\Dm r}\\&
+ \Big(\lambda^3 c_{3,1} \phi\ab{0}{\Dm}^3 e^{-3\Dm r} + \ldots \Big) \delta \phi\ab{0}{\Dp} e^{-\Dp r}\,.
\end{split}
\ee
where the last term is the non-local and unwanted term. However, upon substitution of \eqref{eq:deltaphinonlocal} and all the other radial expansions in the total variation \eqref{eq:deltaStot3}, we find that all the terms which multiply $\delta \phi\ab{0}{\Dp}$ conspire to give zero in the $r \to \infty$ limit. The proper source is therefore still $\phi\ab{0}{\Dm}$, precisely because of the extra boundary terms involving the conjugate momentum. (Notice that also in AlAdS spacetimes there is an interesting relation between the counterterms and the variational principle, see \cite{Papadimitriou:2005ii,Papadimitriou:2010as}.)

For the variation of the total action \eqref{eq:deltaStot3} we then again obtain a simple and finite expression:
\be
\label{eq:deltaStot3source}
\int d^d x  (\Dm - \Dp) (\phi\ab{0}{\Dp} + \phi\ab{1}{\Dp} + \phi\ab{2}{\Dp} + \phi\ab{3}{\Dp}) \delta \phi\ab{0}{\Dm}\,.
\ee
The renormalized one-point function up to this order therefore becomes:
\be
\label{eq:vev5pt}
\vev{\op} = (\Dm - \Dp) (\phi\ab{0}{\Dp} + \phi\ab{1}{\Dp} + \phi\ab{2}{\Dp} + \phi\ab{3}{\Dp})\,.
\ee
Just as at lower orders in $\lambda$, this again reflects the known result that the one-point function to all orders in the sources should be given by $(\Dm - \Dp)$ times $\phi\bs{\Dp}$, at least up to contact terms which may arise in the presence of logarithmic divergences.

Finally, we expect that the general structure exhibited here will persist to higher orders in $\lambda$. In particular, we will at a certain order encounter triple-trace counterterms and so on. However we also expect that we will never need any counterterms linear in the conjugate momentum, since such counterterms cannot be matched to any local counterterms in the field theory.  
\section{Coupling to gravity}
\label{sec:gravity}
In this section we consider the holographic renormalization of a free massive scalar field in a general AlAdS background metric. We suppose that the background metric satisfies the vacuum Einstein equations with a negative cosmological constant but we leave it otherwise unspecified. In particular, in our viewpoint the bulk metric is always dynamical and the boundary metric is therefore kept arbitrary. We are then effectively renormalizing the partition function to all orders in the boundary metric and consequently we are renormalizing correlation functions with an arbitrary number of insertions of the energy-momentum tensor. We will in contrast only be working to second order in the scalar field sources, so our results apply only to correlation functions with at most two scalar operators. Notice that, just like in the previous section, we will not compute any of these correlation functions exactly but rather consider their (holographic) renormalization properties only.

With this setup we expect to capture the following generic class of divergences. In a conformal field theory the renormalization of the two-point function of a scalar operator $\op$ is intricately linked with the singular terms in the $\op \op$ operator product expansion, since these are precisely the terms which need to be regularized and renormalized in order to make the correlation function well-defined also at contact points. Now, for \emph{any} operator $\op$ with a non-zero scaling dimension we should encounter the energy-momentum tensor in the $\op \op$ OPE, since the associated three-point function has to be non-vanishing, and furthermore the coefficient is singular precisely for irrelevant operators. The generality of this statement makes it worthwhile to investigate the associated (holographic) renormalization process in some detail.

In this section we will demonstrate that the holographic renormalization procedure for an arbitrary AlAdS background will be more involved than in the previous section where the background was just AdS in Poincar\'e coordinates. In particular, we will demonstrate that it becomes necessary to take into account the first-order backreaction. Once this is properly done the counterterms take the expected form: they are either completely local functionals of the boundary fields themselves or at least quadratic in the conjugate momentum. The latter counterterms are again interpreted as corresponding to multi-trace counterterms in the field theory.

\subsection{Setup}
We work in Euclidean signature. The bare action for our system is:
\be
S = \frac{1}{2\k^2} \int d^{d+1}x \sqrt{G}(-R + 2 \Lambda) + \hf\int d^{d+1}x \sqrt{G} (\del_\m \Phi \del^\m \Phi + m^2 \Phi^2) - \frac{1}{\k^2} \int d^d x \sqrt{\g} K \,,
\ee
where $\k^2 = 8 \pi G_N$. Notice that we again use $\g_{ij}$ to denote the metric on slices of constant $r$. Our conventions for the curvatures are:
\be
R_{\m \n \r}^{\phantom{\m \n \r}\s} = \del_\n \G_{\m \r}^\s + \G_{\m \r}^{\lambda}\Gamma_{\n\lambda}^\s - (\m \leftrightarrow \n), \qquad \qquad R_{\m \r} = R_{\m \s \r}^{\phantom{\m \s \r}\s}\,.
\ee
The equations of motion are given by:
\be
\begin{split}
&R_{\m \n} - \hf R G_{\m \n} + \Lambda G_{\m \n} = 2 \k^2 T_{\m \n}\\
&T_{\m \n} = \hf \del_\m \Phi \del_\n \Phi - \qt G_{\m \n} (\del_\r \Phi \del^\r \Phi + m^2 \Phi^2)\\
&\square_G \Phi - m^2 \Phi = 0\,.
\end{split}
\ee
We will henceforth set the AdS radius to $1$, which amounts to setting $\Lambda = - d (d-1)/2$. We work in Gaussian normal coordinates with respect to slices of constant radial coordinate $r$. The metric therefore takes the form:
\be
G_{\m \n} dx^\m dx^\n = dr^2 + \g_{ij} dx^i dx^j
\ee
and the extrinsic curvature of a slice of constant $r$ is in our conventions given by:
\be
\label{eq:Kij}
K_{ij} = \hf \dot \g_{ij}\,,
\ee
where the dot denotes a radial derivative. For later convenience we also introduce:
\be
\Pi = \dot \Phi
\ee
and
\be
\D = \hf(d + \sqrt{4 m^2 + d^2}) \qquad \qquad \D(\D- d) = m^2\,.
\ee
The equations of motion can now be rewritten as:
\be
\label{eq:eomfgform}
\begin{split}
\del_r K^i_j + K K^i_j - R^i_j[\g] - \d_j^i \Big(d - \k^2 \frac{\D(\D-d)}{d-1} \Phi^2\Big)+ \k^2 \del_j \Phi \del^i \Phi &=0\\
K^j_i K^i_j + R[\g] - K^2 + d(d-1) + \k^2 (\Pi^2 - \del_k \Phi \del^k \Phi - \D (\D-d)\Phi^2) &= 0\\
\cdel_j K^j_i - \cdel_i K - \k^2 \Pi \del_i \Phi &= 0\\
\dot \Pi + K \Pi + \square_\g \Phi - \D(\D-d)\Phi &= 0\,,
\end{split}   
\ee
where covariant derivatives, curvatures $R_{ij}$ and raised indices are all defined using the $d$-dimensional metric $\g_{ij}$.

We regard the scalar field as a small perturbation on top of a background metric $G_{\m \n}$ which satisfies the vacuum Einstein equations. In the next subsection we first discuss the asymptotic form of the background solution and the corresponding holographic renormalization. Afterwards we consider the scalar field fluctuation and the backreaction onto the metric.

\subsection{Holographic renormalization for Einstein gravity}
As we mentioned in the introduction to this section, we take the background metric $G_{\m \n}$ to be of AlAdS form. This implies that $\g_{ij}$ near the conformal boundary $r \to \infty$ takes the form:
\be
\label{eq:leadingfgterm}
\g_{ij} = e^{2r} g_{(0)ij} + \ldots
\ee
with $g_{(0)ij}$ a non-degenerate boundary metric which sources the boundary energy-momentum tensor. (See \cite{Skenderis:2009kd} for a brief review on AlAdS spacetimes.) Also, here and below the dots denote terms which are subleading in the large $r$ limit. For pure Einstein gravity we find the radial expansion to be:
\be
\g_{ij} = e^{2r}\big(g_{(0)ij} + e^{-2r} g_{(2)ij} + \ldots + e^{-dr} (r \tilde g_{(d)ij} + g_{(d)ij}) + \ldots \big)\,.
\ee
As for the scalar field, the terms $g_{(2)ij}$ until $g_{(d-2)ij}$ as well as $\tilde g_{(d)ij}$ are local functions of the source $g_{(0)ij}$. Actually, in odd boundary dimensions as well as in $d=2$ one finds that $\tilde g_{(d)ij} = 0$. The term $g_{(d)ij}$ is similar to $\phi\bs{\Dp}$ in the scalar field theory example, since it is fixed only by demanding regularity in the interior and then it is non-locally determined in terms of the source $g_{(0)ij}$. It will again define the non-local part of the one-point function of the dual operator, so in this case of the energy-momentum tensor. 
 
Upon substitution of the asymptotic solution into the bare on-shell Einstein-Hilbert action (plus the Gibbons-Hawking boundary term) one finds divergences which need renormalization. As explained in \cite{Papadimitriou:2004ap}, this renormalization is more conveniently done at the level of the first-order variation of the action rather than at the level of the action itself. To illustrate this, consider the first variation of the on-shell gravity action:
\be
\label{eq:varEH}
\delta S = \frac{1}{2\k^2} \int \sqrt{\g} (K^{ij} - K \g^{ij}) \delta \g_{ij} \,,
\ee
where in the Fefferman-Graham coordinate the extrinsic curvature is given in \eqref{eq:Kij}. Notice that \eqref{eq:leadingfgterm} implies that:
\be
\label{eq:leadingKterm}
K_{ij} = \g_{ij} + \ldots
\ee 
Subsituting this into \eqref{eq:varEH} we find that the leading term has a divergence of order $\exp(d r)$. We also immediately see that we can subtract this divergence by defining a counterterm which satisfies:
\be
\label{eq:deltaSctlo}
\delta S_{\text{ct}} = - \frac{1}{2\k^2} \int d^d x \sqrt{\g} (1-d) \g^{ij} \delta \g_{ij} + \ldots
\ee
since adding this variation precisely cancels the leading divergence in \eqref{eq:varEH}. Although one may integrate \eqref{eq:deltaSctlo} back to find the leading counterterm (it is given by $\frac{1}{\k^2} (d-1) \int d^d x \sqrt{\g}$), in practice this integration is never necessary since the one-point function in the presence of sources contains all the necessary information to compute higher-point correlation functions.

To compute the subleading counterterms for the background solution we follow the procedure of \cite{Papadimitriou:2004ap}. We begin by writing out the vacuum Einstein equations which we assumed are satisfied by our solution. They are obtained by setting $\Phi = 0$ in \eqref{eq:eomfgform} and take the form:
\be
\label{eq:vacuumeinstein}
\begin{split}
\del_r K^i_j + K K^i_j - R^i_j[\g] - d \d^i_j &=0\\
\cdel_j K^j_i - \cdel_i K &= 0\\
K^j_i K^i_j + R[\g] - K^2 + d(d-1) &= 0\,.
\end{split}  
\ee
The radial derivative acting on $K^i_j$ can be written in terms of a functional derivative, since at least asymptotically all the radial dependence of $K^i_j$ resides in its functional dependence on the metric $\g_{ij}$. We may therefore use:
\be
\label{eq:delrfunctional}
\del_r K^i_j[\g_{ij}] = \int d^d x \, \dot \g_{kl} \fdel{\g_{kl}} K^i_j = 2 \int d^d x\, K_{kl} \fdel{\g_{kl}} K^i_j \,.
\ee
Upon substitution of \eqref{eq:leadingKterm} we then see that up to subleading terms the radial derivative is equal to the \emph{dilatation operator} which is defined as:
\be
\delta_D \equiv 2 \int d^d x \, \g_{ij} \fdel{\g_{ij}}\,.
\ee
Now, rather than organizing the divergences in eigenfunctions of the radial derivative (\ie powers of $\exp(r)$), it is more convenient to organize them in eigenfunctions of the dilatation operator since the latter is a covariant operator on the slices of constant $r$. We therefore expand:
\be
\label{eq:expandK}
K_i^j = K_{(0)i}^j + K_{(2)i}^j + K_{(4)i}^j + \ldots\,,
\ee
where by definition:
\be
\delta_D K_{(s)i}^j = - s K_{(s)i}^j\,.
\ee
(We assume here an absence of logarithmic terms, see \cite{Papadimitriou:2004ap} for these cases.) To compute the various terms in \eqref{eq:expandK} we use the equations of motion \eqref{eq:vacuumeinstein} but with the radial derivative rewritten as a functional derivative using \eqref{eq:delrfunctional}. Upon substitution of the expansion \eqref{eq:expandK} and collecting terms of equal dilatation weight we can recursively determine the various $K_{(s)i}^j$. To zeroth order we for example find that:
\be
\begin{split}
\delta_D K^i_{(0)j} + K_{(0)} K^i_{(0)j} - d \d_i^j =0\\
\cdel_j K^j_{(0)i} - \cdel_i K_{(0)} = 0\\
K^j_{(0)i} K^i_{(0)j} - K_{(0)}^2 + d(d-1) = 0
\end{split}  
\ee
and these equations are indeed satisfied for 
\be
\label{eq:K0ij}
K_{(0)j}^i = \delta^i_j\,,
\ee
which is in agreement with the leading-order behavior we already found in \eqref{eq:leadingKterm}. At the first subleading order we then find that the equations reduce to:
\be
\begin{split}
\delta_D K^i_{(2)j} + d K^i_{(2)j}+ K_{(2)} \d^i_{(0)j} - R^i_j[\g] &=0\\
\cdel_j K^j_{(2)i} - \cdel_i K_{(2)} &= 0\\
2 K_{(2)} + R[\g] - 2 d K_{(2)} &= 0\,,
\end{split}
\ee
where we used \eqref{eq:K0ij}. These equations then determine:
\be
\label{eq:K2}
K_{(2)} = \frac{R[\g]}{2(d-1)} \qquad \quad (d-2)K^i_{(2)j} = R^i_j[\g] - \frac{R[\g] \d^i_j}{2(d-1)}  \qquad \quad \cdel_j K^j_{(2)i} = \frac{\cdel_i R[\g] }{2(d-1)} \,.
\ee
Notice that for $d \neq 2$ we find that $K_{(2)j}^i$ is completely determined whereas for $d = 2$ the middle equation is trivially satisfied since for a two-dimensional metric $R_{ij} = \hf R \g_{ij}$. 
More generally, this iterative procedure determines all the $K_{(s)j}^i$ for $0 \leq s < d$ but will fail to completely determine $K_{(d)j}^i$. This is the covariant analogue of the fact that $g_{(d)ij}$ is left undetermined by the asymptotic analysis. However the trace and divergence of $K_{(d)j}^i$ are always locally determined, just as in \eqref{eq:K2} for $K_{(2)j}^i$ in $d=2$. These will eventually lead to the diffeomorphism and conformal Ward identities in the dual field theory.

Finally, at the second subleading order we may also determine:
\be
\label{eq:K4}
K_{(4)} = \frac{1}{2(d-1)} (K_{(2)j}^i K_{(2)i}^j - K_{(2)}^2)
\ee
and we will not need the other components of $K_{(4)j}^i$ in what follows so we refer to \cite{Papadimitriou:2004ap} for the exact expression.

Let us now turn to the counterterm action. It is defined implicitly by the formula:
\be
\label{eq:varSct}
\delta S_\text{ct} = - \frac{1}{2\k^2} \int d^d x \sqrt{\g} \sum_{0 \leq s < d} (K_{(s)}^{ij} - K_{(s)}\g^{ij}) \delta \g_{ij}
\ee
and as we mentioned above we do not need to compute it explicitly. Rather, it suffices to note that since all the terms $K_{(s)ij}$ with $0 \leq s < d$ are locally determined functions of $\g_{ij}$, this procedure indeed leads to a local and covariant counterterm action. (In $d=2$ there is an extra logarithmic counterterm which is a topological invariant and therefore cannot be determined in this way. It however does not enter in the one-point function either.)

From \eqref{eq:varEH} and \eqref{eq:varSct} we see directly that the first variation of the renormalized action takes the form: 
\be
\label{eq:deltaSEHren}
\delta (S + S_\text{ct}) = \frac{1}{2\k^2} \int d^d x \sqrt{\g} (K^{ij}_{(d)} - K_{(d)} \g^{ij} + \ldots) \delta \g_{ij}
\ee
where, by using the asymptotic relation between the dilatation operator and the radial derivative, we find that the radial expansion of $K_{(d)ij}= \g_{ik} K_{(d)j}^k$ begins with a term of order $\exp((2-d)r)$. In fact, an explicit computation reveals that \cite{deHaro:2000xn}:
\be
\label{eq:Kdij}
K_{(d)ij} = e^{(2-d)r} (-  \frac{d}{2} g_{(d)ij} + (\text{local}) + \ldots) \,, 
\ee
where we indicated with (local) a certain local function of $g_{(0)ij}$ which appears at the same order in the radial expansion and the dots indicate again subleading terms which vanish as $r \to \infty$. Upon substitution of \eqref{eq:leadingfgterm} and \eqref{eq:Kdij} in \eqref{eq:deltaSEHren} one then finds that the first variation of the renormalized on-shell action is precisely finite,
\be
\lim_{r \to \infty} \delta (S+ S_\text{ct}) = - \frac{d}{4 \k^2} \int d^d x \sqrt{g_{(0)}} (g_{(d)}^{ij} + (\text{local}) ) \delta g_{(0)ij}\,. 
\ee
The one-point function of the energy-momentum tensor is then given by:
\be
\vev{T^{ij}} = - \frac{2}{\sqrt{g_{(0)}}} \fdel{g_{(0)ij}} S = \frac{d}{2 \k^2} g_{(d)}^{ij} + (\text{local})\,.
\ee
By virtue of the equations satisfied for the trace and divergence of $K_{(d)ij}$, one finds that $\vev{T^{ij}}$ satisfies precisely the expected diffeomorphism and conformal Ward identities \cite{Papadimitriou:2004ap}.

\subsection{Scalar field solution}
Let us now consider the holographic renormalization for a free massive scalar field $\Phi$ in the above AlAdS background. To determine the divergences we will follow the same functional approach as in the previous subsection.

The first step is to consider the equation of motion satisfied by the scalar field $\Phi$. Since we regard $\Phi$ as a small perturbation it satisfies the Klein-Gordon equation on a fixed background. The Klein-Gordon equation in our coordinate system was given as the last equation in \eqref{eq:eomfgform} and we repeat here that it takes the form:
\be
\label{eq:eomscalarfgform}
\dot \Pi + K \Pi + \square_\g \Phi - \D(\D-d)\Phi = 0\,,
\ee
where $K$ is the trace of the extrinsic curvature and we recall that $\Pi = \dot \Phi$. Just as for the extrinsic curvature $K_i^j$ in the previous subsection, we may again observe that $\Pi$ at a slice of constant $r$ should be completely determined in terms of the boundary fields $\Phi$ and $\g_{ij}$. Its radial derivative is therefore now given by:
\be
\label{eq:radialderpi}
\del_r \Pi = \int d^d x \,2 K_{ij} \fdel{\g_{ij}} \Pi + \int d^d x \,\Pi \fdel{\Phi} \Pi\,.
\ee
As in section \ref{sec:nologs}, the asymptotic solution to the equation of motion is easily found to be:
\be
\Phi = e^{(\D - d)r} (\phi_{(0)} + \ldots + e^{-(2\D - d) r} \phi_{(2\D - d)} + \ldots)\,.
\ee
and therefore asymptotically:
\be
\Pi = (\D - d) \Phi + \ldots
\ee
The radial derivative is then again seen to be asymptotically equal to the dilatation operator $\delta_D$ which is now defined as:
\be
\label{eq:deltaDphi}
\delta_D  = 2 \int d^d x \,\g_{ij} \fdel{\g_{ij}} + (\D - d) \int d^d x \,\Phi \fdel{\Phi}\,. 
\ee
We again expand $\Pi$ in terms of eigenfunctions of the dilatation operator:
\be
\label{eq:exppi}
\Pi = \Pi_{(0)} + \Pi_{(2)} + \Pi_{(4)} + \ldots\,,
\ee
where by definition
\be
\label{eq:deltaDpis}
\delta_D \Pi_{(s)} = (\D - d - s) \Pi_{(s)}\,.
\ee
Notice the extra shift of $\D - d$ between the label and the actual eigenvalue; this notation is not conventional but will be convenient in what follows.

Plugging now \eqref{eq:radialderpi}, \eqref{eq:exppi} and \eqref{eq:expandK} into \eqref{eq:eomscalarfgform} and using \eqref{eq:deltaDpis} we may iteratively determine $\Pi_{(s)}$ for $s < 2 \D - d$, just as for the coefficients $K_{(s)j}^i$ of the previous subsection. We find:
\be
\begin{split}
\Pi_{(0)} &= (\D - d)\Phi\\
\Pi_{(2)} &= \frac{-1}{2\D - d -2} (\square_\g \Phi + (\D - d) K_{(2)} \Phi)\\
\Pi_{(4)} &= \frac{-1}{2\D - d -4} \Big((\D -d)\Phi K_{(4)} + 2 \int d^d x K_{(2)ij} \fdel[\Pi_{(2)}]{\g_{ij}} \\& \qquad - \frac{\square_\g \Pi_{(2)}}{2\D - d -2} + \frac{3\D - d - 4}{2\D - d -2} K_{(2)} \Pi_{(2)}\Big)\,.
\end{split}
\ee
Substituting \eqref{eq:K2} for $K_{(2)}$ and \eqref{eq:K4} for $K_{(4)}$ we can work out the last expression. To this end we also need:
\be
\label{eq:deltaricci}
\begin{split}
\delta_\g R_{ij}[\g] &= \hf (\cdel^k \cdel_i \delta \g_{jk} + \cdel^k \cdel_j \delta \g_{ik} - \cdel^k \cdel_k \delta \g_{ij} - \cdel_i \cdel_j \g^{kl} \delta \g_{kl})\\
\delta_\g \square_\g \Phi &= - (\delta \g_{ij}) \cdel^i \cdel^j \Phi - \g^{kl}(\cdel_k \delta \g_{lm} - \hf \cdel_m \delta \g_{kl}) \cdel^m \Phi
\end{split}  
\ee
and we eventually find:
\begin{multline}
\label{eq:pi4}
\Pi_{(4)} = \frac{-1}{2\D - d - 4} \Big(
\frac{\D - d}{2(d-1)} \Phi K_{(2)j}^i K_{(2)i}^j - \frac{\D-d}{8(d-1)^3} \Phi R^2
\\ + \frac{1}{(2\D - d -2)^2}
\Big[
\square_\g^2 \Phi
+ \frac{\D - d}{2(d-1)} \square_\g (R \Phi) 
-\frac{(3\D - d -4)(\D - d +1)}{2(d-1)} R\square_\g \Phi\\
+ \frac{2\D - d -2}{2(d-1)} \cdel_k R \cdel^k \Phi + 2 (2\D - d -2) K_{(2)}^{ij} \cdel_i \cdel_j \Phi 
\Big]
\Big)\,.
\end{multline}
For $d > 2$ we may replace $K_{(2)ij}$ in \eqref{eq:pi4} with the solution to the middle equation in \eqref{eq:K2}. However we will focus on $d=2$ below. In this case $K_{(2)ij}$ is not locally determined in terms of $\g_{ij}$ and therefore we kept $K_{(2)ij}$ arbitrary in equation \eqref{eq:pi4}. We did use the other equations in \eqref{eq:K2} which determine its divergence and trace also for $d=2$.

One may continue this expansion and find that all the conjugate momenta $\Pi_{(s)}$ for $s < 2 \D -d$ are again locally determined in terms of $\Phi$. However these terms generically also depend on $K_{(s-2)ij}$ and for $s \geq d + 2$ they are therefore non-locally determined in terms of $\g_{ij}$. We will not work out the detailed form of any more subleading terms here, since the general procedure of dealing with such non-local divergences can be seen already at the level of $\Pi_{(4)}$ by choosing $d=2$.

The on-shell action for the scalar field takes the simple form:
\be
S = \hf \int d^d x \sqrt{\g} \Pi \Phi\,.
\ee
By power-counting (and using the asymptotic relation between the radial derivative and the dilatation operator) The divergent pieces in the on-shell action are given by:
\be
S_\text{div} = \hf \int d^d x \sqrt{\g} \sum_{0 \leq s < 2\D - d}\Pi_{(s)} \Phi\,.
\ee
One would be tempted to define (minus) this as the proper counterterm action, since it is covariant and a local function of the field $\Phi$. However this is not the correct counterterm action because we have not yet incorporated the backreaction onto the metric. In the next subsection we will explain why the backreaction necessarily has to be taken into account.

\subsection{The first-order backreaction}
In this subsection we consider the first-order backreaction onto the metric. We choose a radial-axial gauge for the metric variation, so we set $\delta G_{rr} = \delta G_{ri} = 0$. We then write:
\be
\g_{ij} \to \g_{ij} + \delta \g_{ij} = \g_{ij} + \k^2 \s_{ij}\,,
\ee
where the factor of $\k^2$ is inserted for later convenience. From the linearization of the equations of motion \eqref{eq:eomfgform} we obtain that $\s_{ij}$ is quadratic in $\Phi$ and zeroth order in $\k^2$.

Consider now the variation in the on-shell renormalized action as a result of this change in the metric. Since for \emph{any} first-order variation the bulk term vanishes by the equation of motion, the resulting change is precisely the boundary term given in \eqref{eq:deltaSEHren} with $\d \g_{ij} = \k^2 \s_{ij}$, so:
\be
\label{eq:deltaSEHrens}
\frac{1}{2} \int d^d x \sqrt{\g} (K^{ij}_{(d)} - K_{(d)} \g^{ij} + K^{ij}_{(d+2)} - K_{(d + 2)}\g^{ij} + \ldots) \s_{ij}\,.
\ee
This is a boundary term and therefore depends crucially on the asymptotic behavior of $\s_{ij}$. As in our scalar field example of section \ref{sec:nologs}, the correct boundary condition for $\s_{ij}$ is dictated by imposing that the boundary metric does not change, so by imposing that the term of order $\exp(2r)$ in the radial expansion $\s_{ij}$ vanishes. We therefore write:
\be
\label{eq:expsij}
\s_{ij} = \ldots + 0\, e^{2r} + \ldots
\ee
If the term of order $\exp(2r)$ were the leading term in $\s_{ij}$ then we could ignore the backreaction, since the other terms in \eqref{eq:deltaSEHrens} conspire to be of total order $\exp(-2r)$ to leading order. However, from the second equation in \eqref{eq:eomfgform} one may deduce that the leading term in $\s_{ij}$ has to be of order $\exp(2(\D - d)r)$ and therefore more divergent than $\exp(2r)$ if $\D > d$. For irrelevant operators there thus have to be leading terms in \eqref{eq:expsij} and substitution into \eqref{eq:deltaSEHrens} will then lead to new divergences if $K^{ij}_{(d)} - K_{(d)} \g^{ij}$ is non-zero.

The \emph{total} bare action quadratic in the scalar field is given by:
\be
\label{eq:Sbaretotal}
S_\text{bare} = \frac{1}{2} \int d^d x \sqrt{\g} (K^{ij}_{(d)} - K_{(d)} \g^{ij} + \ldots) \s_{ij}
+ \hf \int d^d x \sqrt{\g} \Pi \Phi \,.
\ee
Our perturbative expansion is in terms of the number of sources and in terms of Newton's constant. However the first term in \eqref{eq:Sbaretotal} is of precisely the same order as the second as both terms are second order in the number of sources and zeroth order in Newton's constant. We conclude that if the first term does not vanish as $r \to \infty$, so if the dual operator is irrelevant \emph{and} the one-point function of the energy-momentum tensor is non-zero, then the backreaction onto the metric cannot be ignored. 

In the holographic renormalization literature the incorporation of the backreaction has been investigated before. For example in \cite{deHaro:2000xn} the authors considered the backreaction for marginal and relevant operators and demonstrated how the Ward identities related to the energy-momentum tensor receive the corrections which are expected by the presence of a scalar source. In \cite{Bianchi:2001de,Bianchi:2001kw} the holographic renormalization was performed by solving the non-linear field equations but it was observed that the backreaction for relevant operators could be ignored in certain cases where the equations asymptotically decouple. Finally, in \cite{Papadimitriou:2004rz} the linearized equations did not decouple and in such cases one necessarily has to take into account the backreaction, even at the linearized level and for relevant operators.  

\subsection{Computation of the backreaction}
In this section we compute the first-order backreaction of the scalar field on the metric. We henceforth set $\k^2 = 1$ but the dependence on $\k^2$ can be trivially reinstated. We denote the variation of $K_{ij}$ by $\lambda_{ij}$, so:
\be
K_{ij} \to K_{ij} + \lambda_{ij} \qquad \qquad \lambda_{ij} = \hf \dot \s_{ij}\,.
\ee
Notice that this implies that:
\be
\begin{split}
K_i^j &\to K_i^j + \lambda_i^j - \s^j_k K^k_i\\ 
K &\to K + \lambda - \s^j_i K^i_j\\
\del_r K_i^j &\to \del_r K_i^j + \del_r \lambda_i^j - \s^j_k \del_r K^k_i + 2 K_i^k \s_k^l K_l^j - 2 K_i^k \lambda_k^j\,,
\end{split} 
\ee
where $\lambda_i^j = \g^{jk} \lambda_{kj}$, $\lambda = \gamma^{ij}\lambda_{ij}$ and similarly all other indices will be raised using $\g_{ij}$. We also write:
\be
R_{ij}[\g] \to R_{ij}[\g] + X_{ij}\,,
\ee
where the linearized Ricci tensor $X_{ij} = \delta R_{ij}$ was already given in \eqref{eq:deltaricci}, of course now with $\delta \g_{ij} = \s_{ij}$. We also write $\s = \g^{ij} \s_{ij}$ and $X = \g^{ij} X_{ij}$. With these notations the linearization of the relevant equations of motion \eqref{eq:eomfgform} becomes:
\bea
&- 2 \s_i^k K^i_j K^j_k + 2 \lambda_i^j K_j^i - 2 \lambda K + X - \s^i_j R^j_i + 2 K \s_i^j K_j^i + \Pi^2 - \del_k \Phi \del^k \Phi - \D (\D - d) \Phi^2 = 0\nonumber \\
&\del_r \lambda_i^j - d \s_i^j + 2 K_i^k \s_k^l K_l^j - 2 K_i^k \lambda_k^j + (\lambda - \sigma_k^l K_l^k) K_i^j  \label{eq:lineomfgform} \\ &+ K \lambda_i^j - X_i^j + \del_i \Phi \del^j \Phi + \delta_i^j \frac{\D(\D -d)}{d-1} \Phi^2  = 0\nonumber
\eea
and we also need that:
\be
\label{eq:deflambda}
\lambda_i^j = \hf \del_r \s_i^j + \s_i^k K_k^j \,.
\ee
Just as for $K_{ij}$ and $\Pi$, it will be most convenient to compute the backreaction $\s_{ij}$ as a function of $\g_{ij}$ and $\Phi$ since this allows us to directly obtain a covariant expression. The radial derivative on $\s_i^j$ is then rewritten as:
\be
\label{eq:delrsigma}
\del_r \s_i^j = \Big(\int d^d x 2 K_{kl} \fdel{\g_{kl}} + \int d^d x \Pi \fdel{\Phi}\Big) \s_i^j
\ee
and similarly for $\lambda_i^j$. We again organize the solution in eigenfunctions of the dilatation operator which takes the same form as in \eqref{eq:deltaDphi}. We write:
\be
\label{eq:expansionsigmalambda}
\begin{split}
\s_i^j &= \s_{(0)i}^j + \s_{(2)i}^j + \s_{(4)i}^j + \ldots\\
\lambda_i^j &= \lambda_{(0)i}^j + \lambda_{(2)i}^j + \lambda_{(4)i}^j + \ldots
\end{split}
\ee
where by definition:
\be
\delta_D \s_{(s)i}^j = (2 \D - 2 d - s) \s_{(s)i}^j \,.
\ee
It is then tedious but straightforward to work out the coefficients. To this end one first replaces the radial derivative in both \eqref{eq:lineomfgform} and in \eqref{eq:deflambda} with the operator defined in \eqref{eq:delrsigma}. The next step is to substitute the expansions \eqref{eq:expansionsigmalambda} for $\s_i^j$ and $\lambda_i^j$ as well as the expansions \eqref{eq:expandK} and \eqref{eq:exppi} for $K_i^j$ and $\Pi$. Finally one collects the terms of equal dilatation weight and solves for the various coefficients. In this way one obtains the traces:
\be
\s_{(0)} = \frac{-d}{2(d-1)}\Phi^2 \qquad \qquad \lambda_{(0)} = \frac{-d(\D - d + 1)}{2(d-1)}\Phi^2\,,
\ee
from which one subsequently obtains that:
\be
\s_{(0)i}^j = \frac{-1}{2(d-1)} \Phi^2 \delta_i^j \qquad \qquad \lambda_{(0)i}^j = \frac{-(\D - d+ 1)}{2(d-1)}\Phi^2 \delta_i^j \,.
\ee
At the first subleading order one finds again first the traces:
\be
\begin{split}
\s_{(2)} &= \frac{1}{2(d-1)(\D - d -1)(2\D - d -2)}\Big(- 2 \Phi\square_\g\Phi + \frac{d-2}{2(d-1)} R[\g] \Phi^2\Big)\\
\lambda_{(2)} &= \frac{1}{2(d-1)(\D - d -1)(2\D - d -2)}\Big(d \Phi \square_\g \Phi +\frac{(d-2)(\D - d -1)}{2(d-1)} R[\g] \Phi^2\Big)\\& \qquad  + (\D - d)\s_{(2)}\,,
\end{split} 
\ee
and the full coefficients are then given by:
\be
\begin{split}
\lambda_{(2)i}^j &= \frac{1}{2(d-1)(2\D - d -2)} \Big( \Phi \square_\g \Phi + \frac{\D-d}{2(d-1)} R[\g] \Phi^2 \Big) \delta_i^j \\&\qquad + (\D - d) \s_{(2)i}^j - \frac{1}{2(d-1)} \Phi^2 K_{(2)i}^j\\
\sigma_{(2)i}^j &= \frac{-1}{(\D - d - 1)(2 \D - d - 2)} \Big(
\frac{d^2 - d \D - d + 2}{2(d-1)}K_{(2)i}^j \Phi^2
+ \cdel_i \Phi \cdel^j \Phi \\&\qquad - \frac{d-2}{4(d-1)} \cdel_i \cdel^j \Phi^2
+ \frac{\d_i^j}{2(d-1)} \Big[
\frac{\D - d}{2(d-1)} R[\g] \Phi^2
+ 2 \Phi \square_\g \Phi
- \hf \square_\g \Phi^2
\Big] 
\Big)\,.
\end{split}
\ee
Notice that we again kept $K_{(2)ij}$ explicit in these expressions as we will shortly focus again on the case where $d=2$. In principle one may continue this computation of the backreaction to any given order.

\subsection{Renormalization including the backreaction}
Let us now plug our asymptotic solution into the on-shell action \eqref{eq:Sbaretotal}. We again specialize to $d=2$ but for future reference we keep $d$ explicit in many of the expressions below. Up to the order we are interested in here we find that:
\be
S = \frac{1}{2} \int d^d x \sqrt{\g} (K_{(2)}^{ij} - K_{(2)} \g^{ij} + K_{(4)}^{ij} - K_{(4)} \g^{ij}) \s_{ij} + \hf \int d^d x \sqrt{\g} \Pi \Phi\,. 
\ee
Notice that we included the subleading terms involving $K_{(4)ij}$. Normally such terms would vanish in the limit $r \to \infty$ but since $\s_{ij}$ is more divergent than the background metric $\g_{ij}$ these terms actually do contribute. Upon substitution of the various expansions we find that:
\be
S = S_{(0)} + S_{(2)} + S_{(4)} + \ldots\,,
\ee 
where the leading divergence is not modified by the backreaction:
\be
S_{(0)} = \hf \int d^d x \sqrt{\g} (\D - d) \Phi^2\,,
\ee
but the first subleading term is now:
\be
\begin{split}
S_{(2)} &= \hf \int d^d x \sqrt{\g} [\Pi_{(2)} \Phi + (K_{(2)}^{ij} - K_{(2)} \g^{ij})\s^{ij}_{(0)}]\\
&= -\hf  \int d^d x \sqrt{\g}\Big(\frac{\Phi \square_\g \Phi}{2\D - d - 2} + \frac{(2\D - 2d - 1)}{4(d-1)(2\D - d -2)} R[\g] \Phi^2\Big)
\end{split}
\ee
and the second subleading piece becomes:
\be
\label{eq:S4}
S_{(4)} = \hf \int d^d x \sqrt{\g} [\Pi_{(4)} \Phi + (K_{(4)}^{ij} - K_{(4)} \g^{ij})\s^{ij}_{(0)} + (K_{(2)}^{ij} - K_{(2)} \g^{ij})\s^{ij}_{(2)}]\,.
\ee
Before writing out the full expression, let us look at the divergences \emph{linear} in $K_{(2)ij}$ which by the arguments of section \ref{sec:nologs} would be non-local in $d=2$. Since its trace and divergence are locally determined even in $d=2$, see equation \eqref{eq:K2}, the only possible non-local divergence is the term of the form:
\be
K^{ij}_{(2)} \Phi \cdel_i \cdel_j \Phi\,,
\ee 
which in $d=2$ cannot, even after integration by parts, be written as a local function of the metric $\g_{ij}$. From \eqref{eq:pi4} one finds that the corresponding term in the scalar field part of \eqref{eq:S4} has the coefficient:
\be
\Pi_{(4)} \Phi  = \ldots + \frac{-2}{(2\D - d - 4)(2\D - d -2)} K^{ij}_{(2)} \Phi \cdel_i \cdel_j \Phi + \ldots\,,
\ee
whereas in the gravity part of the action it enters via the term $K_{(2)}^{ij} \s_{(2)ij}$. After an integration by parts one finds that:
\be
\int d^d x \sqrt{\g} K_{(2)}^{ij} \s_{(2)ij} = \int d^d x \sqrt{\g} (\ldots + \frac{1}{(\D - d -1)(2\D - d - 2)} K^{ij}_{(2)} \Phi \cdel_i \cdel_j \Phi  + \ldots)\,.
\ee
Adding the coefficients one finds:
\begin{align}
&S_{(4)} = \ldots + \hf \int d^d x \sqrt{\g} K^{ij}_{(2)} \Phi \cdel_i \cdel_j \Phi \Big(
\frac{-2}{(2\D - d - 4)(2\D - d - 2)} + \frac{1}{(\D - d - 1)(2\D - d - 2)}
\Big) + \ldots \nonumber \\
&= \ldots + \hf \int d^d x \sqrt{\g} K^{ij}_{(2)} \Phi \cdel_i \cdel_j \Phi \Big(
\frac{d -2}{(2\D - d - 2)(\D - d - 1)(2\D - d - 4)}
\Big) + \ldots\,,
\end{align}
so our potentially non-local divergence vanishes \emph{precisely} when $d=2$. We may again refer to this as a pseudo-non-local divergence: although it may arise by power-counting its coefficient in the counterterm action actually vanishes. This cancellation should not be accidental but rather is a reflection in the bulk of the renormalizability of the dual CFT. We emphasize that the observed cancellation is however the result of rather lengthy computation and it would be desirable to understand it on a more structural level.

The full expression for the divergences at this order reads:
\be
\label{eq:uglydiv}
\begin{split}
&S_{(4)} = \hf \int d^d x \sqrt{\g} \Big( \frac{-1}{(2\D -d-4)(2\D-d-2)^2} \Phi\square_\g \Phi\\
&- \frac{12 - 28\D + 22\D^2 - 6 \D^3 + 20d - 34 \D d + 14\D^2 d +13d^2 - 10\D d + 2 d^3}{(d-1)(\D-d-1)(2\D-d-4)(2\D-d-2)^2} \Phi R \square_\g \Phi\\
&- \frac{d-2}{4(d-1)(2\D -d -2)(\D - d- 1)(2\D - d- 4)} R \cdel_k \Phi \cdel^k \Phi\\
&-\frac{(d-2)(d^2 -3\D d + d + 2 \D^2 -4)}{16(2\D-d-4)(d-1)^3(\D - d-1)(2\D-d-2)} R^2 \Phi^2\\
&+\frac{8+8\D-8\D^2-14d+ 4\D d+6 \D^2 d+5d^2-9\D d^2+3d^3}{4(\D-d-1)(2\D-d-2)(2\D-d-4)(d-1)} K_{(2)i}^j K_{(2)j}^i \Phi
\Big)\,.
\end{split}
\ee
We may define the counterterm action as minus $S_{(0)} + S_{(2)} + S_{(4)}$. Notice again the presence of a multi-trace counterterm proportional to $K_{(2)i}^j K_{(2)j}^i$ on the last line of \eqref{eq:uglydiv}.

Finally by power-counting we find that only the leading counterterm has the right powers of $\exp(r)$ to contribute to the one-point function. (Recall that we used a similar argument for the counterterms involving boxes in the previous section.) Therefore the one-point function is of the standard form:
\be
\delta S = (\Dm - \Dp) \int d^d x \sqrt{g_{(0)}} \phi\ab{0}{\Dp}\delta \phi\ab{0}{\Dm}\,,
\ee
just as for the scalar field theory example of section \ref{sec:nologs}. We would like to emphasize that in the presence of logarithmic divergences this result is augmented by various contact terms. We will report on this in \cite{me}.

\section{Conclusions}
We have analyzed in two toy model examples the general features of holographic renormalization in the presence of sources for irrelevant operators. The structure we have found extends the standard holographic renormalization results. Namely, as we already summarized in the introduction to this paper, we have found pseudo-non-local divergences, multi-trace counterterms and the need to take the backreaction into account even at the level of scalar two-point functions.

We expect that our results are much more generally valid. For example, by general OPE arguments we expect multi-trace counterterms to arise already in the four-point function of operators with sufficiently high difference between their scaling dimensions. Either way it would be interesting to obtain a more general understanding of the structures we have identified, in particular we would like to have a general proof of the absence of any counterterms which are linear in the conjugate momentum. This would be equivalent to a proof that the dual field theory can be renormalized when working to arbitrary finite order in the sources for irrelevant operators.

In string theory compactifications most operator dimensions are such that logarithmic divergences will arise in the renormalization procedure. We will therefore in \cite{me} extend the results of this paper to include such cases as well and derive the associated anomalous conformal Ward identities and operator mixing between single- and multi-trace operators.

Recently in for example \cite{Guica:2008mu} the notion of the \emph{asymptotic symmetry group} has been revived as a tool in the analysis of holography for non-AlAdS spacetimes. In those cases one performs a similar analysis as in \cite{Brown:1986nw}, where the asymptotic symmetry group was used to correctly compute the central charge of a CFT whose energy-momentum tensor sector is described by Einstein gravity in AdS$_3$. As was reviewed in \cite{Skenderis:2009kd}, for AlAdS spacetimes this analysis has been embedded into a more precise holographic dictionary which extends beyond just Einstein gravity. In particular, within this framework the precise falloff conditions for normalizable modes can be \emph{computed} from the AlAdS hypothesis and the equations of motion. Finiteness of the charges is guaranteed by the holographic renormalization procedure. To see how the computation of these falloff conditions is affected by irrelevant deformations one may consider a small `normalizable' fluctuation $\delta \Phi$ of the scalar field $\Phi$ of section \ref{sec:nologs}, so a fluctuation where $\delta \phi\bs{\Dm} = 0$. The correct boundary conditions for $\delta \Phi$ are then directly seen to be a function not just of $\delta \Phi$ but also of the conjugate momentum $\del_r \delta \Phi$. It would be interesting to further investigate this issue and its possible repercussions on the asymptotic symmetry group analysis for non-AlAdS spacetimes.

\section*{Acknowledgments}
We would like to thank Nikolay Bobev, Leonardo Rastelli, Shlomo Razamat and Marika Taylor for useful discussions. We would especially like to thank Kostas Skenderis for many valuable discussions and his comments on the draft of this paper.

\appendix

\section{The bulk-to-bulk propagator}
\label{app:bulkbulk}
The asymptotic expansion of the solution to the equations of motion in section \ref{sec:nologs} is not easily recovered from more familiar expressions. For example, in the literature it is customary to use the bulk-bulk propagator in position space, see for example the review \cite{D'Hoker:2002aw}. In our coordinate system this bulk-bulk propagator is written as:
\be
\label{eq:bulkbulkhyperg}
K (x^\m,x'{}^\m) = C \x^{\Dp} F(\frac{\Dp}{2},\frac{\Dp + 1}{2}; \frac{\Dp - \Dm}{2} + 1; \x^2 )
\ee
where $x^\m = (r , x^i)$ and:
\be
\x = \frac{2 e^{- r - r'}}{e^{-2r} + e^{-2r'} + (x - x'{})^i(x - x')_i}
\ee
and the normalization is given by:
\be
C = \frac{\Gamma(\Dp)}{2^{\Dp} \pi^{(\Dp + \Dm)/2} (\Dp - \Dm)\G(\hf(\Dp - \Dm))}
\ee
The function $K(x^\m,x'{}^\m)$ is symmetric in its arguments, satisfies the equation:
\be
(\square_G - m^2) K(x^\m,x'{}^\m) = \frac{1}{\sqrt{G}} \delta^{d+1}(x^\m - x'{}^\m)
\ee
and has the normalizable boundary condition:
\be
\label{eq:limitbulkbulk}
K(x^\m,x'{}^\m) = O( e^{-\D_+ r} ) \qquad \text{as} \qquad r \to \infty \,,
\ee 
where it is understood that $r'$ is kept finite in the limit. Using this bulk-bulk propagator, the first-order correction to the free-field solution is then tentatively written as:
\be
\label{eq:phi1bulkbulk}
\Phi\as{1}(x^\m) = \lambda \int d^{d+1}x' \sqrt{G(x'{}^\m)}\, K(x^\m,x'{}^\m) \Phi\as{0}^2(x'{}^\m)\,.
\ee
Applying the naive limit \eqref{eq:limitbulkbulk} however does not reproduce the asymptotic behavior of section \ref{sec:nologs}. The reason for this is that the above integral does not converge, which can be easily seen by using the asymptotic expansions of the integrand:
\be
\sqrt{G(x'{}^\m)}\, K(x^\m,x'{}^\m) \Phi\as{0}^2(x'{}^\m) \sim e^{-\D_- r'} \qquad \text{as} \qquad r' \to \infty 
\ee
which is indeed divergent for irrelevant operators, even for finite $r$. We therefore cannot trust the expansion inside the integral.

One method to deal with the divergence in \eqref{eq:phi1bulkbulk} is to impose a cutoff at a large but finite $r_0$. This method renders everything manifestly finite but it modifies the solution $\Phi\as{1}$. One may then substitute the solution with a cutoff into the on-shell action and try to cancel the divergences that arise as one sends $r_0 \to \infty$ with counterterms. It would be interesting to compare this prescription with the one we use in the main text, but we stress that the philosophy employed in this method is rather different. Namely, in principle there is nothing wrong with the solution $\Phi\as{1}$ which is by definition just the first-order correction to the solution to the equations of motion. Rather it is our method of computing $\Phi\as{1}$, namely using \eqref{eq:phi1bulkbulk}, which fails (essentially because $\Phi\as{1}$ cannot satisfy normalizable boundary conditions). For this reason the divergence in \eqref{eq:phi1bulkbulk} does not mean that $\Phi\as{1}$ itself needs any form of regularization and renormalization as is the case for the the on-shell action. When one cuts off the integral in \eqref{eq:phi1bulkbulk} one appears to unnecessarily modify $\Phi\as{1}$ such that the equation of motion is no longer completely satisfied. Whether or not this presents insurmountable difficulties for the holographic renormalization procedure remains to be seen.

A second method of dealing with the divergence in \eqref{eq:phi1bulkbulk} is by analytic continuation. One may for example compute the integral for a value of $\D$ where it converges and then analytically continue $\D$ to the desired value. By construction the expression for $\Phi\as{1}$ so obtained will satisfy the correct equation of motion. This procedure is technically more involved than the previous one but we will show below that it indeed leads to the correct asymptotic expansion presented in section \ref{sec:nologs}.

To exhibit the leading behaviour of $\Phi\as{1}$ we will first Fourier transform along the boundary directions. For the free-field solution in Poincar\'e coordinates this leads to the familiar expression:
\be
\label{eq:phi0solved}
\Phi\as{0} = \int \frac{d^d k}{(2\pi)^d} \phi_{(0)}(k_i) \frac{|k|^h}{2^{h - 1} \G(h)} e^{-d r /2} K_h (|k| e^{-r}) e^{i k_j x^j}
\ee
where $|k| = \sqrt{k_j k^j}$ and we defined:
\be
h \equiv \hf(\D_+ - \D_-) = \hf \sqrt{ d^2 + 4 m^2}.
\ee
Indeed, upon substitution of the expansion of the Bessel function:
\be
\label{eq:Besselexpansion}
K_h(z) =  \G(h) 2^{h - 1} z^{-h} ( 1 + \frac{z^2}{4(1-h)} + \ldots) +
 \G(-h) 2^{-h -1} z^h ( 1+ \frac{z^2}{4(1+h)} + \ldots)
\ee
we directly obtain the behavior of $\Phi\as{0}$ as given in \eqref{eq:phias0}.

Let us now similarly rewrite the bulk-bulk propagator in Fourier space. To this end we notice that the \emph{inhomogeneous} Klein-Gordon equation:
\be
\square_G \Phi - m^2 \Phi = \lambda \Phi
\ee
has the solutions:
\be
e^{-d r/2} K_{i \m}(|k| e^{-r}) \qquad \qquad e^{-d r/2} I_{i \m}(|k| e^{-r}) 
\ee
where $i \m = \sqrt{h^2 + \lambda}$. The second of these solutions blows up in the interior, so as $r \to - \infty$, and therefore does not satisfy the boundary conditions there. Furthermore, if we insist that the solution be normalizable as $r \to \infty$ we need $\m$ to be real. With these boundary conditions the spectrum of allowed eigenvalues is $\lambda \in (-\infty,- h^2)$ or $\m \in \Real^+$. The solutions for different values of $\m$ are orthogonal, more precisely we find:
\be
\int dr K_{i\m}(|k|e^{-r}) K_{i\m'}(|k|e^{-r}) = \frac{\pi^2}{2\m \sinh(\pi \m)} (\delta (\m - \m') + \delta( \m + \m') )
\ee
(The integral may be regulated by inserting an extra $e^{-\a r}$ in the integrand and the desired result is then obtained by taking the limit $\a \downarrow 0$.) By the standard arguments of Sturm-Liouville theory these modes form a complete set:
\be
e^{-dr} \delta(r -r') = \int_0^\infty d\m\frac{2}{\pi^2} \m \sinh(\pi \m) e^{-d r/2} K_{i\m}(|k|e^{-r}) e^{-dr'/2} K_{i\m}(|k|e^{-r'})
\ee
The bulk-bulk propagator in Fourier space is then given by:
\be
\label{eq:bulkbulkfour}
\begin{split}
&K(x^\m, x'{}^\m) =\\  &\int \frac{d^d k}{(2\pi)^d}  \int_{-\infty}^\infty d\m \frac{- \m \sinh(\pi \m)}{\pi^2 (\m^2 + h^2)} e^{i k_i x^i -d r/2} K_{i\m}(|k|e^{-r}) e^{i k_i x'{}^i-dr'/2} K_{i\m}(|k|e^{-r'})
\end{split}
\ee
In this equation, we used the manifest symmetry $\m \leftrightarrow - \m$ to extend the integral over all $\m \in \Real$. To verify that it also satisfies the right boundary conditions one substitutes the asymptotic expansion of the first Bessel function and keeps $r'$ finite. This leads to:
\be
\begin{split}
&K(x^\m, x'{}^\m) =\\  &\int \frac{d^d k}{(2\pi)^d}  \int_{-\infty}^\infty d\m \frac{- \m \sinh(\pi \m)}{\pi^2 (\m^2 + h^2)} e^{i k_i x^i -d r/2} e^{- i k_i x'{}^i-dr'/2} K_{i\m}(|k|e^{-r'}) \\ &\times
\Big(\G(i\m) 2^{i\m -1} |k|^{-i\m} e^{i \m r} + \ldots + \G(-i\m) 2^{-i\m -1} |k|^{i\m} e^{-i \m r}+ \ldots \Big) 
\end{split}
\ee
The integral splits into two parts which can be evaluated by contour deformation in the complex $\m$ plane. (For all finite $|k| \exp(-r')$ the second Bessel function is an analytic function of $\mu$. Its large $\mu$ asymptotics may furthermore be regularized by integrating against a source as we will demonstrate explicitly below.) Since $r > 0$ the first integral converges in the upper half plane and the second integral in the lower half plane. We pick up the poles at $\m = \pm i h$ and find:
\be
K(x^\m,x'{}^\m) = -\frac{e^{-\Dp r}}{2h} \int \frac{d^d k}{(2\pi)^d} e^{i k_i (x - x'{})^i} \frac{|k|^h}{2^{h-1} \G(h)} e^{-dr'/2} K_{h}(|k|e^{-r'})
\ee
so we recover indeed the requested boundary condition \eqref{eq:limitbulkbulk}. Upon comparison with \eqref{eq:phi0solved} we furthermore recover the familiar property that the bulk-bulk propagator asymptotes precisely to the bulk-boundary propagator times a factor $-\exp(-\Dp r)/(2h)$, see \cite{D'Hoker:2002aw}.

Of course, upon subtitution of this form of the bulk-bulk propagator in equations like \eqref{eq:phi1bulkbulk} we encounter precisely the same divergences as before and we need to analytically continue in $\D$ to make the integral finite. As we will shortly see, the expression \eqref{eq:bulkbulkfour} is much more amenable to this approach than \eqref{eq:bulkbulkhyperg}. Let us for example recover the leading behavior of $\Phi\as{1}$. To this end we substitute the expression:
\be
\Phi\as{0} = \phi\ab{0}{\Dm} \exp(-\Dm r)
\ee
into \eqref{eq:phi1bulkbulk}. (Although for non-constant $\phi\ab{0}{\Dm}$ this $\Phi\as{0}$ does not solve the free equation of motion, this ansatz suffices to illustrate the leading asymptotic behavior of $\Phi\as{1}$.) After interchanging the integrals over $r'$ and $\m$ we find:
\be
\Phi\as{1} = \lambda \int \frac{d^d k}{(2\pi)^d} \int d\m \frac{-\m \sinh(\pi \mu)}{\pi^2(\m^2 + h^2)} e^{ik_i x^i - d r/2} K_{i\m}(|k|e^{-r}) (\phi\ab{0}{\Dm} * \phi\ab{0}{\Dm}) C[k_i,\m]\,,
\label{eq:phi1int}
\ee
where we defined the convolution operator:
\be
\phi\ab{0}{\Dm} * \phi\ab{0}{\Dm} \equiv \int \frac{d^d q}{(2\pi)^d} \phi\ab{0}{\Dm}(q) \phi\ab{0}{\Dm}(k-q) 
\ee
and:
\be
\label{eq:Cdef}
C[k_i,\m] = \int dr' e^{- (3\Dm - \Dp) r'/2} K_{i\m}(|k|e^{-r'})
\ee
If $\m$ is real then \eqref{eq:Cdef} converges for $3 \Dm - \Dp > 0$, so for $\D < 3d /4$, which is generically not the case. We may however obtain a finite result by analytic continuation of $\D$ to the region of convergence and obtain:
\be
\label{eq:C}
C[k_i,\m] = 2^{-2+(\Dp - 3\Dm)/2} |k|^{(\Dp - 3\Dm)/2} \Gamma(\qt(3\Dm -\Dp - 2 i \m)) \Gamma(\qt(3\Dm -\Dp + 2 i \m))
\ee
We now substitute \eqref{eq:C} and the asymptotic expansion of the Bessel function \eqref{eq:Besselexpansion} into \eqref{eq:phi1int}, split the integral over $\m$ into two parts and evaluate each part by contour deformation. The difference with the previous case is that the gamma functions in \eqref{eq:C} give rise to extra poles at:
\be
\m = \pm \frac{i}{2} (3\Dm - \Dp + 4 n) \qquad \qquad n \in \{0,1,2,\ldots\}
\ee
Remembering that we assumed that $3 \Dm - \Dp > 0$, we find the contributions from the leading poles to conspire precisely to:
\be
\Phi\as{1} = \lambda f(2\Dm) \phi\ab{0}{\Dm}^2  e^{-2\Dm r} + \ldots
\ee
We may now analytically continue back $\D$ to its original value to recover precisely the leading term $\phi\ab{1}{2\Dm}$ in \eqref{eq:expphi1}. The subleading terms can be obtained in a similar matter. This shows that the asymptotics in \eqref{eq:expphi1} can indeed be recovered from the full solution, albeit in a rather non-trivial way.

\bibliographystyle{utphys}
\bibliography{biblio}

\end{document}